\documentclass[a4paper,UKenglish,cleveref, autoref, thm-restate]{lipics-v2021}

\nolinenumbers
\usepackage[ruled,vlined, linesnumbered ]{algorithm2e}

\title{Sidestepping Barriers for Dominating Set in Parameterized Complexity}

\author{Ioannis Koutis}{New Jersey Institute of Technology, USA}{ikoutis@njit.edu}{}{}
\author{Michał Włodarczyk}{University of Warsaw, Poland}{m.wlodarczyk@mimuw.edu.pl}{}{}
\author{Meirav Zehavi}{Ben-Gurion University of the Negev, Israel}{meiravze@bgu.ac.il}{}{}

\authorrunning{I. Koutis, M. Wlodarczyk and M. Zehavi}

\Copyright{Ioannis Koutis, Michał Włodarczyk and Meirav Zehavi}

\ccsdesc[500]{Theory of computation~Parameterized complexity and exact algorithms}

\keywords{Dominating Set, Parameterized Complexity, Approximation Algorithms} 

\funding{Supported by BGU–NJIT Joint Seed Research Fund and ERC Starting Grant titled PARAPATH.}

\EventEditors{John Q. Open and Joan R. Access}
\EventNoEds{2}
\EventLongTitle{42nd Conference on Very Important Topics (CVIT 2016)}
\EventShortTitle{CVIT 2016}
\EventAcronym{CVIT}
\EventYear{2016}
\EventDate{December 24--27, 2016}
\EventLocation{Little Whinging, United Kingdom}
\EventLogo{}
\SeriesVolume{42}
\ArticleNo{23}
\begin{document}
\maketitle

\begin{abstract}
We study the classic {\sc Dominating Set} problem with respect to several prominent parameters. Specifically, we present algorithmic results that sidestep time complexity barriers by the incorporation of either approximation or larger parameterization. Our results span several parameterization regimes, including: (i,ii,iii) time/ratio-tradeoff for the parameters {\em treewidth}, {\em vertex modulator to constant treewidth} and {\em solution size}; (iv,v) FPT-algorithms for the parameters {\em vertex cover number} and {\em feedback edge set number}; and (vi) compression for the parameter {\em feedback edge set number}.
\end{abstract}

\section{Introduction}

The {\sc Dominating Set} problem is one of the most central problems in Parameterized Complexity~\cite{downey.fellows,cyganetal.book}. The input to {\sc Dominating Set} consists of an $n$-vertex graph $G$, and the objective is to output a minimum-size subset $U\subseteq V(G)$  that is a {\em dominating set}---that is, the closed neighborhood of $U$ in $G$ equals $V(G)$, or, in other words, every vertex in $V(G)\setminus U$ is adjacent in $G$ to at least one vertex in $U$. When parameterized by the solution size and stated as a decision problem, the input also consists of a non-negative integer $k$, and the objective is to determine whether there exists a subset $U\subseteq V(G)$ of size at most $k$ that is a  dominating set.

From the perspective of Parameterized Complexity, {\sc Dominating Set} parameterized by the sought solution size $k$ is very hard. First, {\sc Dominating Set} is W[2]-complete~\cite{downey.fellows} (and, clearly, in XP).\footnote{We refer to Section \ref{sec:prelims} for notations and concepts not defined in the introduction.} In fact, {\sc Dominating Set} and {\sc Set Cover} are the two most well-studied W[2]-hard problems in Parameterized Complexity. Moreover, under the Strong Exponential Time Hypothesis (SETH), {\sc Dominating Set} cannot be solved in $f(k)\cdot n^{k-\epsilon}$ time~\cite{patrascu.2010}. Still, for every integer $k\geq 7$, {\sc Dominating Set} is solvable in $n^{k+o(1)}$ time~\cite{eisenbrand_complexity_2004}.

From the perspective of approximation (and parameterized approximation), {\sc Dominating Set} is also very hard. Unless P=NP, {\sc Dominating Set} does not admit a polynomial-time $(1-\epsilon)\ln n$-approximation algorithm for any fixed $\epsilon>0$~\cite{DBLP:conf/stoc/DinurS14} (see also~\cite{feige.threshold.98}). However, it admits a polynomial-time $(\ln n - \ln\ln n +O(1))$-approximation algorithm \cite{vazirani}. Moreover, under the SETH, {\sc Dominating Set} does not even admit a $g(k)$-approximation $f(k)\cdot n^{k-\epsilon}$-time algorithm for any computable functions $g$ and $f$ of $k$ and fixed $\epsilon>0$~\cite{karthik.parameterized.2019}. (Observe that this statement generalizes the one above by \cite{patrascu.2010}.) Similar (but weaker) results of hardness of approximation in the setting of Parameterized Complexity also exist under other assumptions, including the ETH, W[1]$\neq$FPT, and the $k$-SUM hypothesis \cite{karthik.parameterized.2019}.

Concerning structural parameters, the most well-studied parameters in Parameterized Complexity are treewidth and vertex cover number~\cite{downey.fellows,cyganetal.book}.  Regarding {\sc Dominating Set}, on the positive side, the problem is easily solvable in  $O(3^{\mathsf{tw}}\cdot n)$ time~\cite{van_rooij_dynamic_2009}. Here, $\mathsf{tw}$ is the treewidth of the input graph. However, under the SETH, {\sc Dominating Set} cannot be solved in $(3-\epsilon)^{\mathsf{tw}}\cdot n^{O(1)}$ time for any fixed $\epsilon>0$~\cite{lokshtanov.known.2018}. Similarly to the case of the parameter solution size $k$, we again have (essentially) matching upper and lower bounds in terms of time complexity. Moreover, it is not hard to see that, under any of the SETH and the Set Cover Conjecture, {\sc Dominating Set} cannot be solved in $(2-\epsilon)^{\mathsf{vc}}\cdot n^{O(1)}$ time for any fixed $\epsilon>0$ (see Section \ref{sec:vc}). Here, $\mathsf{vc}$ is the vertex cover number of the input graph.

Lastly, we note that the weighted version of {\sc Dominating Set} is, similarly, approximable in polynomial time within factor $O(\log n)$, and solvable in  $O(3^{\mathsf{tw}}\cdot n)$ time.  Further, being more general, all negative results carry to it as well.   
    
\subsection{Our Contribution} 
Our contribution is fivefold, concerning five different parameterizations. 

\subparagraph*{I. Treewidth.} First, in Section~\ref{sec:tw}, we consider the treewidth $\mathsf{tw}$ of the given graph as the parameter.  We prove that  {\sc Weighted Dominating Set}  admits a 2-approximation $O(\sqrt{6}^{\mathsf{tw}}\cdot\mathsf{tw}^{O(1)}\cdot n)$-time algorithm.  Our proof is based on ``decoupling'' the task of domination of the entire input graph $G$ into two separate tasks: we compute a partition $(V_1,V_2)$ of $G$ based on a proposition of \cite{DBLP:conf/soda/LokshtanovMRSZ21}, and then consider the domination of each $V_i$ separately. We remark that the way that we use the aforementioned proposition is very different than the way it is originally used in~\cite{DBLP:conf/soda/LokshtanovMRSZ21}. Here, we remind that under the SETH, {\sc Dominating Set} cannot be solved in $(3-\epsilon)^{\mathsf{tw}}\cdot n^{O(1)}$ time for any fixed $\epsilon>0$.

\subparagraph*{II. Modulator to Constant Treewidth.} Second, in Section~\ref{sec:consttw}, we consider the parameter $\mathsf{tw}_d$, the minimum-size of a vertex modulator of the given graph to treewidth $d$, for any fixed $d\in\mathbb{N}$. We note that, for any graph $G$, $\mathsf{tw}\leq \mathsf{tw}_d+d$. We prove that  {\sc Weighted Dominating Set}  admits a $2$-approximation $O(2^{\mathsf{tw}_d}\cdot n)$-time algorithm. As before, our proof is based on ``decoupling'' the task of domination of the entire input graph $G$ into two separate tasks: now, these are the task of the domination of the modulator, and the task of the domination of the rest of $G$. Unlike before, the resolution of these two tasks is different.  Concerning the tightness of our result, we refer the reader to Conjecture~\ref{conj:constTwLower}, where we conjecture that the same time complexity cannot be attained by an exact algorithm.

\subparagraph*{III. Vertex Cover Number.} Third, in Section~\ref{sec:vc}, we consider the vertex cover number $\mathsf{vc}$ of the given graph as the parameter.  We prove that  {\sc Weighted Dominating Set}  admits an $O(2^{\mathsf{vc}}\cdot n)$-time algorithm.  Our proof is partially based on the idea of our algorithm for the parameter $\mathsf{tw}_d$, combined with the observation that some vertices in the independent set (being the complement of the given vertex cover) are ``forced'' to be picked, after having chosen which vertices to pick from the vertex cover. From the perspective of impossibility results, we observe that under any of the SETH and the Set Cover Conjecture, {\sc Dominating Set} cannot be solved in $(2-\epsilon)^{\mathsf{vc}}\cdot n^{O(1)}$ time for any fixed $\epsilon>0$, thus our time complexity is tight. %Due to space constraints, these results are deferred to Appendix~\ref{sec:vc}.

\subparagraph*{IV. Feedback Edge Set Number.} Fourth, in Sections~\ref{sec:fes} and~\ref{sec:compression_}, we consider the parameter $\mathsf{fes}$ (feedback edge set number), the minimum-size of a set of edges whose removal transforms the input graph into a forest. Notice that this parameter is a relaxation of $\mathsf{tw}_d$ (for any fixed $d\in\mathbb{N}$), which, in turn, is a relaxation of $\mathsf{tw}$. We present two theorems. The first theorem is that {\sc Dominating Set} admits an $O(3^{\frac{\mathsf{fes}}{2}}\cdot n)$-time algorithm. To this end, we prove the following combinatorial lemma, which is of independent interest: For any graph $G$, $\mathsf{tw}_2(G)\leq \frac{\mathsf{fes}(G)}{2}$.
(Moreover, we prove that there exists an algorithm that, given a graph $G$, outputs a subset $M\subseteq V(G)$ such that $|M|\leq \frac{\mathsf{fes}(G)}{2}$ and $\mathsf{tw}(G-M)\leq 2$ in $O(\mathsf{fes}(G)+n)$ time.)
The second theorem is that an instance $G$ of the {\sc Dominating Set} problem with $\mathsf{fes}(G)=k$ can be compressed in linear time into a ``relaxed'' instance of the problem on a graph $\hat{G}$ with $O(k)$ edges, requiring the minimum domination of a subset of the vertices in $\hat{G}$. %Due to space constraints, the proof of this last statement is deferred to Appendix~\ref{sec:compression_}.

\subparagraph*{V. Solution Size.} Fifth, we consider the solution size $k$ as the parameter. We prove that, for any fixed $0\leq \alpha<1$, {\sc Dominating Set} admits a  $((1-\alpha)\ln n + O(1))$-approximation $n^{\alpha k+O(1)}$-time algorithm. The proof of this theorem is the simplest one in our article, based on the combination of an exhaustive search (to uncover part of the solution) and a known approximation algorithm. This approach somewhat resembles that of \cite{DBLP:journals/ipl/CyganKW09}. Here, we remind that (under plausible complexity-theoretic assumptions) it is unlikely for an exact algorithm to run in $n^{\alpha k+O(1)}$ time, or for a polynomial-time algorithm to have approximation factor $((1-\alpha)\ln n + O(1))$. %Due to space constraints, this result is deferred to Appendix~\ref{sec:solsize}. 

\subsection{Other Related Works} 

Here, we briefly survey a few works not already mentioned that are directly relevant or related to ours. First, we note that {\sc Dominating Set} can be solved in linear time on series-parallel graphs~\cite{Kikuno1983ALA}. In particular, the class of series-parallel graphs is a (strict) subclass of the class of graphs of treewidth~$2$.

The restriction of {\sc Dominating Set} to planar graphs is known to be both solvable in  $2^{O(\sqrt{k})}n$ time (thus, it is in FPT) and admit an EPTAS~\cite{fomin.dominating.2006}. Similar results exist also for more general graph classes, such as $H$-minor free graphs \cite{fomin.minor.free} and graphs of bounded expansion~\cite{amiri2018}. Moreover, {\sc Dominating Set} is also in FPT (specifically, it is solvable in $2^{O(k)}n^{O(1)}$ time, and admits a polynomial kernel) on claw-free graphs, but it remains W[2]-complete on $K_{1,t}$-free graphs for any $t\geq 4$ \cite{cygan_dominating_2011,DBLP:conf/icalp/HermelinMLW11}. Additionally, {\sc Dominating Set} is W[1]-hard on unit disk graphs~\cite{marx_parameterized_2006}.
    
With respect to exact exponential-time algorithms, the currently best known running time upper bound is of $O(1.4969^n)$, based on the branch-and-reduce method \cite{vanroo}.

\section{Preliminaries}\label{sec:prelims}

Given a function $f: U\rightarrow \mathbb{R}$ and a subset $U'\subseteq U$, let $f(U')=\sum_{u\in U'}f(u)$. When $f$ is interpreted as a weight function, we will refer to $f(U')$ as the weight of $U'$.

\subparagraph*{Standard Graph Notation.} Throughout the article, we deal with simple, finite, undirected graphs. Given a graph $G$, let $V(G)$ and $E(G)$ denote its vertex and edge sets, respectively. When no confusion arises, we denote $|V(G)|=n$ and $|E(G)|=m$. Given a vertex $v\in V(G)$, let $N_G(v)=\{u\in V(G): \{u,v\}\in E(G)\}$ and $N_G[v]=N_G(v)\cup\{v\}$. Given a subset $U\subseteq V(G)$, let $N_G(U)=\bigcup_{u\in U}N_G(u)\setminus U$ and $N_G[U]=\bigcup_{u\in U}N_G[u]$. When $G$ is clear from context, we drop it from the subscripts. Given a subset $U\subseteq V(G)$, let $G[U]$ denote the subgraph of $G$ induced by $U$, and let $G-U$ denote the graph $G[V(G)\setminus U]$.  Given a subset $F\subseteq E(G)$, let $G-F$ denote the graph on vertex set $V(G)$ and edge set $E(G)\setminus F$.
Given subsets $A,B\subseteq V(G)$, we say that $A$ {\em dominates} $B$ if for every $b\in B$, $N[b]\cap A\neq\emptyset$. A {\em dominating set} of $G$ is a subset $U\subseteq V(G)$ that dominates $V(G)$.
A {\em vertex cover} of $G$ is a subset $U\subseteq V(G)$ such that $G-U$ is edgeless (i.e., $E(G-U)=\emptyset$). Let $\mathsf{vc}(G)$ denote the minimum size of a vertex cover of $G$.
A {\em feedback edge set} of $G$ is a subset $F\subseteq E(G)$ such that $G-F$ is a forest. Let $\mathsf{fes}(G)$ denote the minimum size of a feedback edge set of $G$. We note that $\mathsf{vc}(G)$ and $\mathsf{fes}(G)$ are incomparable. When $G$ is immaterial or clear from context, we denote $\mathsf{vc}=\mathsf{vc}(G)$ and $\mathsf{fes}=\mathsf{fes}(G)$.
A {\em cactus} is a connected graph in which any two simple cycles have at most one vertex in common. We will slightly abuse this term: given graph $G$ such that each connected component of $G$ is a cactus, we will call $G$ a cactus as well.

\subparagraph*{Problem Definitions.} The {\sc Dominating Set} problem is defined as follows: The input consists of an $n$-vertex graph $G$, and the objective is to output a minimum-size dominating set in $G$. The {\sc Weighted Dominating Set} is defined similarly: Here, the input also consists of a weight function $w: V(G)\rightarrow \mathbb{N}$, and the objective is to output a minimum-weight dominating set in $G$. When parameterized by the solution size $k$, we consider the decision version of {\sc Dominating Set}, and suppose that the input also consists of a non-negative integer $k$. Then, the objective is to determine whether $G$ has a dominating set of size at most $k$. Parameterization by the solution size $k$ can also be defined for {\sc Weighted Dominating Set}. However, we find it to be somewhat less natural (particularly when considered from the perspective of parameterized approximation), and therefore we do not consider it in this paper. Still, we mention that, here, the objective would be to find a minimum-weight dominating set in $G$ among all those of size at most $k$, if one exists.

The {\sc Set Cover} problem is defined as follows: The input consists of a universe $U$ and a family ${\cal F}\subseteq 2^U$ of subsets of $U$ , and the objective is to output a minimum-size subfamily ${\cal S}\subseteq{\cal F}$ such that $U=\bigcup{\cal S}$. (Here, without loss of generality, we suppose that $U=\bigcup{\cal F}$, so there necessarily exists a solution.) The {\sc Weighted Set Cover} problem is defined similarly: Here, the input also consists of a weight function $w:{\cal F}\rightarrow\mathbb{N}$, and the objective is to output a minimum-weight subfamily ${\cal S}\subseteq{\cal F}$ such that $U=\bigcup{\cal S}$. We require a slight generalization of this problem, called {\sc Generalized Weighted Set Cover}, defined as follows: The input $(U,{\cal F},w)$ is the same, and the objective is to output, for every subset $A\subseteq U$, a minimum-weight subfamily ${\cal S}_A\subseteq{\cal F}$ such that $U_A=\bigcup{\cal S}_A$.

\subparagraph*{Width Measures.} The treewidth of a graph is a standard measure of its ``closeness''  to a tree, defined as follows. 

\begin{definition}\label{def:treeDecomp}
A {\em tree decomposition} of a graph $G$ is a pair ${\cal T}=(T,\beta)$, where $T$ is a rooted tree and $\beta$ is a function from $V(T)$ to $2^{V(G)}$, that satisfies the following conditions.
\begin{itemize}
\item For every edge $\{u,v\}\in E(G)$, there exists $x\in V(T)$ such that  $\{u,v\}\subseteq \beta(x)$.
\item For every vertex $v\in V(G)$, $T[\{x\in V(T): v\in\beta(x)\}]$ is a tree on at least one vertex.  
\end{itemize}
The {\em width} of $(T,\beta)$ is $\max_{x\in V(T)} |\beta(x)|-1$. The {\em treewidth} of $G$, denoted by $\mathsf{tw}(G)$, is the minimum width over all tree decompositions of $G$. For every $x\in V(T)$, $\beta(x)$ is called a {\em bag}, and $\gamma(x)$ denotes the union of the bags of $x$ and the descendants of $x$ in $T$. 
\end{definition}

When $G$ is immaterial or clear from context, we denote $\mathsf{tw}=\mathsf{tw}(G)$. Following the standard custom in parameterized algorithmics, when we consider a problem parameterized by $\mathsf{tw}$, we suppose that we are given a tree decomposition $\cal T$ of width $\mathsf{tw}$. Also, following the standard custom, we do not rely on a supposition that the width of $\cal T$ is $\mathsf{tw}$ in the sense that, if the width of $\cal T$ is larger, then our algorithmic result holds where $\mathsf{tw}$ is replaced by this width.

Given a graph $G$, a {\em path decomposition} of $G$ is a tree decomposition $(T,\beta)$ where $T$ is a path, and the {\sc pathwidth} of $G$ is the minimum width over all path decompositions of $G$. 

For $d\in\mathbb{N}\cup\{0\}$ and a graph $G$, let $\mathsf{tw}_d(G)$ denote the minimum size of a vertex set whose deletion from $G$ results in a graph of treewidth at most $d$. Observe that, for any graph $G$, $\mathsf{tw}_0(G)=\mathsf{vc}(G)$. When $G$ is immaterial or clear from context, we denote $\mathsf{tw}_d=\mathsf{tw}_d(G)$. Following the standard custom in parameterized algorithmics, when we consider a problem parameterized by $\mathsf{tw}_d$, we suppose that we are given a vertex set $M$ of size $\mathsf{tw}_d$ whose deletion from the input graph results in a graph of treewidth at most $d$. Also, following the standard custom, we do not rely on a supposition that $|M|=\mathsf{tw}_d$ in the sense that, if $|M|$ is larger, then our algorithmic result holds where $\mathsf{tw}_d$ is replaced by $|M|$.

For the design of algorithms based dynamic programming, it is convenient to work with {\em nice} tree decompositions, defined as follows.
\begin{definition}
A tree decomposition $(T,\beta)$ of a graph $G$ is {\em nice} if for the root $r=\mathsf{root}(T)$ of $T$, $\beta(r)=\emptyset$, and each node $x\in V(T)$ is of one of the following types.
%\vspace{-0.5em}
\begin{itemize}
\item {\bf Leaf}: $x$ is a leaf in $T$ and $\beta(x)=\emptyset$.
\item {\bf Forget}: $x$ has one child, $y$, and there is a vertex $v\in\beta(u)$ such that $\beta(x)=\beta(y)\setminus\{v\}$.
\item {\bf Introduce}: $x$ has one child, $y$, and there is a vertex $v\in\beta(x)$ such that $\beta(x)\setminus\{v\}=\beta(y)$.
\item {\bf Join}: $x$ has two children, $y$ and $z$, and $\beta(x)=\beta(y)=\beta(z)$.
\end{itemize}
\end{definition}

\begin{proposition}[\cite{DBLP:journals/siamcomp/Bodlaender96}]\label{prop:nice}
Given a tree decomposition $(T,\beta)$ of a graph $G$,  a nice tree decomposition of $G$ of the same width as $(T,\beta)$ can be constructed in linear-time (specifically, $O(\mathsf{w}^{O(1)}\cdot n)$ where $\mathsf{w}$ is the width of $(T,\beta)$).
\end{proposition}

Due to Proposition \ref{prop:nice}, when we deal with nice tree decompositions of width $\mathsf{tw}$, we suppose that $|V(T)|\leq  O(\mathsf{tw}^{O(1)}\cdot n)$.

\subparagraph*{Parameterized Complexity.} Let $\Pi$ be an NP-hard problem. In the framework of Parameterized Complexity, each instance of $\Pi$ is associated with a {\em parameter} $k$. Here, the goal is to confine the combinatorial explosion in the running time of an algorithm for $\Pi$ to depend only on $k$. Formally, we say that $\Pi$ is {\em fixed-parameter tractable (FPT)} if any instance $(I, k)$ of $\Pi$ is solvable in $f(k)\cdot |I|^{O(1)}$ time, where $f$ is an arbitrary function of $k$. A weaker request is that for every fixed $k$, the problem $\Pi$ would be solvable in polynomial time. Formally, we say that $\Pi$ is {\em slice-wise polynomial (XP)} if any instance $(I, k)$ of $\Pi$ is solvable in $f(k)\cdot |I|^{g(k)}$ time, where $f$ and $g$ are arbitrary functions of $k$.
Parameterized Complexity also provides methods to show that a problem is unlikely to be FPT. Here, the concept of W-hardness replaces the one of NP-hardness. For more information, we refer the reader to the book \cite{cyganetal.book}.

Essentially tight conditional lower bounds for the running times of parameterized algorithms often rely on the {\em Exponential-Time Hypothesis (ETH)}, the {\em Strong ETH (SETH)} and the {\em Set Cover Conjecture}. To formalize the statements of ETH and SETH, recall that  given a formula $\varphi$ in conjuctive normal form (CNF) with $n$ variables and $m$ clauses, the task of {\sc CNF-SAT} is to decide whether there is a truth assignment to the variables that satisfies $\varphi$. In the {\sc $p$-CNF-SAT} problem, each clause is restricted to have at most $p$ literals. First, ETH asserts that {\sc 3-CNF-SAT} cannot be solved in $O(2^{o(n)})$ time. Second, SETH asserts that for every fixed $\epsilon<1$, there exists a (large) integer $p=p(\epsilon)$ such that {\sc $p$-CNF-SAT} cannot be solved in $O((2-\epsilon)^{n})$ time. Moreover, the Set Cover Conjecture states that for every fixed $\epsilon<1$ {\sc Set Cover} cannot be solved in $O((2-\epsilon)^{n})$ time where $n$ is the size of the universe.

 Let $P$ and $Q$ be two parameterized problems. A {\em compression} (or {\em compression algorithm} for $P$ is a polynomial-time procedure that, given an instance $(x,k)$ of $P$, outputs an equivalent instance $(x',k')$ of $Q$ where $|x'|,k'\leq f(k)$ for some computable function $f$. Then, we say that $P$ admits a {\em compression of size $f(k)$}. When $P=Q$, compression is called {\em kernelization}.

When a problem is parameterized by the solution size $k$, the concept of parameterized approximation must be clarified (given that we then deal with a decision problem). Here, the objective becomes the following where the sought approximation ratio is some $\alpha$: If there exists a solution of size at most $k$, then we seek an $\alpha$-approximate solution; else, we can output any solution. Of course, we do not know (as part of the input) which case is true.

\section{Parameter: Treewidth}\label{sec:tw}

%\section{{\sc Dominating Set} Parameterized by Size of Vertex Modulator to Constant Treewidth}\label{sec:tw}

We will use a translation of {\sc (Weighted) Dominating Set} into two instances of an easier problem, to attain a $2$-approximation algorithm for {\sc (Weighted) Dominating Set}.

\begin{theorem}\label{thm:twAlgo}
The {\sc Weighted Dominating Set} problem parameterized by $\mathsf{tw}$ admits a $2$-approximation $O(\sqrt{6}^{\mathsf{tw}}\cdot\mathsf{tw}^{O(1)}\cdot n)$-time algorithm.
\end{theorem}

Towards the proof of this theorem, we first define the easier problem that we aim to solve.

\begin{definition}[{\bf Half-Width Domination}]
In the {\sc Weighted Half-Width Domination} problem, the input consists of a graph $G$, a vertex-weight function $w: V(G)\rightarrow\mathbb{N}$, a nice tree decomposition ${\cal T}=(T,\beta)$ of $G$ of width $\mathsf{tw}$, and a subset $D\subseteq V(G)$ such that for every $x\in V(T)$, $|\beta(x)\cap D|\leq \frac{\mathsf{tw}}{2}+O(1)$. The objective is to compute a subset $S\subseteq V(G)$ of minimum weight that dominates $D$.
\end{definition}

We would have liked to call the algorithm in the following proposition in order to directly solve the {\sc Weighted Half-Width Domination} problem.

\begin{proposition}[\cite{cyganetal.book}]\label{prop:domSetTwAlgo}
%There exists an algorithm that, given an instance $(G,w)$ of {\sc Weighted Dominating Set} and a tree decomposition ${\cal T}$ of $G$, outputs a minimum-weight dominating set of $G$ in time $O(3^t\cdot n)$ where $t$ is the weight of ${\cal T}$. \hly{cite}
The {\sc Weighted Dominating Set} problem admits an $O(3^\mathsf{tw}\cdot n)$-time algorithm.
\end{proposition}

Unfortunately, we cannot use it in a black-box manner---we need to modify the dynamic programming table used in the proof. Intuitively, we let vertices outside $D$ correspond to two states (chosen, not chosen) instead of three (chosen, not chosen and dominated, not chosen and not dominated). This is done in the following lemma.

\begin{lemma}\label{lem:halfWidth}
The {\sc Weighted Half-Width Domination} problem admits an $O(\sqrt{6}^\mathsf{tw}\cdot\mathsf{tw}^{O(1)}\cdot n)$-time algorithm.
\end{lemma}

\begin{proof}
We first describe the algorithm. Let $(G,w,D,{\cal T}=(T,\beta)$ be the give input. We use dynamic programming, and start with the formal definition of the table, denoted by $\mathfrak{M}$. For every $x\in V(T)$, partition $(X,Y)$ of $\beta(x)\setminus D$ and partition $(\widehat{X},\widehat{Y}_1,\widehat{Y}_2)$ of $\beta(x)\cap D$, we have a table entry $\mathfrak{M}[x,(X,Y),(\widehat{X},\widehat{Y}_1,\widehat{Y}_2)]$. The order of the computation is done by postorder on $T$ (where the order of computation of entries with the same first argument is arbitrary). Then, the basis corresponds to the case where $x$ is a leaf. In this case, we initialize $\mathfrak{M}[x,(\emptyset,\emptyset),(\emptyset,\emptyset,\emptyset)]=0$. Now, suppose that $x$ is not a leaf. Then, we use the following recursive formulas:
\begin{enumerate}
\item In case $x$ is of type Forget, let $y$ be its child and $v\in\beta(y)\setminus\beta(x)$. Then, we compute $\mathfrak{M}[x,(X,Y),(\widehat{X},\widehat{Y}_1,\widehat{Y}_2)]$:
\begin{itemize}
\item If $v\in D$, then we take $\min\{\mathfrak{M}[y,(X,Y),(\widehat{X}\cup\{v\},\widehat{Y}_1,\widehat{Y}_2)],\mathfrak{M}[y,(X,Y),(\widehat{X},\widehat{Y}_1\cup\{v\},\widehat{Y}_2)]\}$.
\item Else, we take $\min\{\mathfrak{M}[y,(X\cup\{v\},Y),(\widehat{X},\widehat{Y}_1,\widehat{Y}_2)],\mathfrak{M}[y,(X,Y\cup\{v\}),(\widehat{X},\widehat{Y}_1,\widehat{Y}_2)]\}$.
\end{itemize}

\item In case $x$ is of type Introduce, let $y$ be its child and $v\in\beta(x)\setminus\beta(y)$. Then, we compute $\mathfrak{M}[x,(X,Y),(\widehat{X},\widehat{Y}_1,\widehat{Y}_2)]$:
\begin{itemize}
\item If $v\in X\cup\widehat{X}$, then we take $\mathfrak{M}[y,(X\setminus\{v\},Y),(\widehat{X}\setminus\{v\},\widehat{Y}_1\setminus N_G(v),\widehat{Y}_2\cup (\widehat{Y}_1\cap N_G(v)))]+w(v)$.
\item Else, if $v\in Y\cup\widehat{Y}_2$, then we take $\mathfrak{M}[y,(X,Y\setminus\{v\}),(\widehat{X},\widehat{Y}_1,\widehat{Y}_2\setminus\{v\})]$.
\item Else, if $v\in N_G(X\cup\widehat{X})$, then we take $\mathfrak{M}[y,(X,Y),(\widehat{X},\widehat{Y}_1\setminus\{v\},\widehat{Y}_2)]$.
\item Else, we take $\infty$.
\end{itemize}

\item In case $x$ is of type Join, let $y$ and $z$ be its children. Then, $\mathfrak{M}[x,(X,Y),(\widehat{X},\widehat{Y}_1,\widehat{Y}_2)]$ equals:
\[\displaystyle{\min_{(Y^y_1,Y^z_1)\ \mathrm{partition\ of} \widehat{Y}_1}\{\mathfrak{M}[y,(X,Y),(\widehat{X},Y^y_1,\widehat{Y}_2\cup(\widehat{Y}_1\setminus Y^y_1))]+\mathfrak{M}[z,(X,Y),(\widehat{X},Y^z_1,\widehat{Y}_2\cup(\widehat{Y}_1\setminus Y^z_1))]\}}\]\[-w(X\cup\widehat{X}).\]
\end{enumerate}

Eventually, the algorithm returns the weight stored in $M[\mathsf{root}(T),(\emptyset,\emptyset),(\emptyset,\emptyset,\emptyset)]$, where the matching itself can be retrieved by backtracking its computation (specifically, collecting the vertices inserted into $X\cup\widehat{X}$).

Because for every $x\in V(T)$, $|\beta(x)\cap D|\leq \frac{\mathsf{tw}}{2}+O(1)$, and $|V(T)|\leq O(\mathsf{tw}^{O(1)}\cdot n)$, we derive that the size of $\mathfrak{M}$ is $O(2^{\frac{\mathsf{tw}}{2}}\cdot 3^{\frac{\mathsf{tw}}{2}}\cdot\mathsf{tw}^{O(1)}\cdot n)=O(\sqrt{6}^{\mathsf{tw}}\cdot\mathsf{tw}^{O(1)}\cdot n)$. So, clearly, the computation of all entries corresponding to leaves, Forget nodes and Introduce nodes can be done within this time bound. The computation of all Join nodes can also be done within this time bound by the use of {\em fast subset convolution} in the exact same manner as it is done for the known exact algorithm for {\sc Weighted Dominating Set} parameterized by $\mathsf{tw}$ (see Section 11.1 in \cite{cyganetal.book}).

Correctness can be proved by straightforward induction on the order of the computation (following the same lines as for the exact algorithm for {\sc Weighted Dominating Set} parameterized by~$\mathsf{tw}$).

\begin{claim}
Every entry $\mathfrak{M}[x,(X,Y),(\widehat{X},\widehat{Y}_1,\widehat{Y}_2)]$ stores the minimum weight of a dominating set $S$ of $G[\gamma(x)]$ that satisfies:
\begin{itemize}
\item $S\cap \beta(x)=X\cup\widehat{X}$.
\item $S$ dominated $\widehat{Y}_1$.
\end{itemize}
If such a matching does not exist, then the entry stores $\infty$.
\end{claim}

This completes the proof.
\end{proof}

For the proof of Theorem \ref{thm:twAlgo}, we also need the following result.

\begin{proposition}[\cite{DBLP:conf/soda/LokshtanovMRSZ21}, Corollary]\label{prop:getHalfWidth}
There exists an $O(n\cdot\mathsf{tw})$-time algorithm that, given a graph $G$ and a tree decomposition ${\cal T}=(T,\beta)$ of $G$ of width $\mathsf{tw}$, outputs a partition $(V_1,V_2)$ of $V(G)$ such that for every $i\in\{1,2\}$ and $x\in V(T)$, $|\beta(x)\cap V_i|\leq \frac{\mathsf{tw}}{2}+O(1)$.
\end{proposition}

We proceed with the following immediate observation.

\begin{observation}\label{obs:mergeProblems}
Let $(G,w)$ be an instance of {\sc Weighted Dominating Set}, and let $D\subseteq V(G)$. Then,
\begin{enumerate}
\item Any dominating set $S\subseteq V(G)$ of $G$ dominates both $D$ and $V(G)\setminus D$
\item Let $S_1\subseteq V(G)$ dominate $D$, and $S_2\subseteq V(G)$ dominate $V(G)\setminus D$. Then, $S_1\cup S_2$ is a dominating set of $G$.
\end{enumerate}
\end{observation}

Now, we are ready to conclude the correctness of Theorem \ref{thm:twAlgo}.

\begin{proof}[Proof of Theorem \ref{thm:twAlgo}]
We first describe the algorithm. Let $(G,w,{\cal T})$ be an instance of {\sc Weighted Dominating Set} parameterized by $\mathsf{tw}$. Due to Proposition \ref{prop:nice}, we can suppose that $\cal T$ is nice. First, we call the algorithm of Proposition \ref{prop:getHalfWidth} with $(G,w,{\cal T})$ as input, and let $(V_1,V_2)$ denote its outputs. Then, we call the algorithm of Lemma \ref{lem:halfWidth} twice, once with $(G,w,{\cal T},V_1)$ as input and once with $(G,w,{\cal T},V_2)$ as input, and let $S_1$ and $S_2$ denote their outputs. We return $S=S_1\cup S_2$.

Clearly, due to Proposition \ref{prop:getHalfWidth} and Lemma \ref{lem:halfWidth}, the algorithm runs in $O(\sqrt{6}^{\mathsf{tw}}\cdot\mathsf{tw}^{O(1)}\cdot n)$ time. For correctness, first note that due to the second item in Observation \ref{obs:mergeProblems}, the output set $S$ is a dominating set of $G$. Moreover, due to the first item in Observation \ref{obs:mergeProblems}, the optimums of $(G,w,{\cal T},V_1)$ and $(G,w,{\cal T},V_1)$ as instances of {\sc Weighted Half-Width Domination} are both bounded from above by the optimum of $(G,w,{\cal T})$ as an instance of {\sc Weighted Dominating Set}. Hence, both of $w(S_1),w(S_2)$ are bounded from above by the optimum of $(G,w,{\cal T})$ as an instance of {\sc Weighted Dominating Set}, which implies that $w(S)$ is bounded from above by twice the optimum of $(G,w,{\cal T})$ as an instance of {\sc Weighted Dominating Set}. This completes the proof.
\end{proof}

\section{Parameter: Size of Vertex Modulator to Constant Treewidth}\label{sec:consttw}

%\section{{\sc Dominating Set} Parameterized by Size of Vertex Modulator to Constant Treewidth}\label{sec:consttw}

Similarly to the proof in Section \ref{sec:tw}, will use a translation of an instance of {\sc (Weighted) Dominating Set} into two instances of two easier problems, to attain a $2$-approximation algorithm for {\sc (Weighted) Dominating Set}. Here, however, the translation is somewhat different.

\begin{theorem}\label{thm:constTwAlgo}
For any fixed constant $d\geq 1$, the {\sc Weighted Dominating Set} problem parameterized by $\mathsf{tw}_d$ admits a $2$-approximation $O(2^{\mathsf{tw}_d}\cdot n)$-time algorithm.
\end{theorem}

Towards the proof of this theorem, we first define the two easier problems that we will aim to solve.

\begin{definition}[{\bf Modulator Domination}]
Let $d\in\mathbb{N}\cup\{0\}$. In the {\sc Weighted $d$-Modulator Domination} problem, the input consists of a graph $G$, a vertex-weight function $w: V(G)\rightarrow\mathbb{N}$, and a subset $M\subseteq V(G)$ such that the treewidth of $G-M$ is at most $d$. The objective is to compute a subset $S\subseteq V(G)$ of minimum weight that dominates $M$.
\end{definition}

\begin{definition}[{\bf Decomposition Domination}]
Let $d\in\mathbb{N}\cup\{0\}$. In the {\sc Weighted $d$-Decomposition Domination} problem, the input consists of a graph $G$, a vertex-weight function $w: V(G)\rightarrow\mathbb{N}$, and a subset $M\subseteq V(G)$ such that the treewidth of $G-M$ is at most $d$.  The objective is to compute a subset $S\subseteq V(G)$ of minimum weight that dominates $V(G)\setminus M$.
\end{definition}

To reuse the result for {\sc Modulator Domination} in Section \ref{sec:vc}, we require a slight generalization of the problem, defined as follows.

\begin{definition}[{\bf Generalized Modulator Domination}]
Let $d\in\mathbb{N}\cup\{0\}$. In the {\sc Generalized Weighted $d$-Modulator Domination} problem, the input consists of a graph $G$, a vertex-weight function $w: V(G)\rightarrow\mathbb{N}$, and a subset $M\subseteq V(G)$ such that the treewidth of $G-M$ is at most $d$. The objective is to compute, for every subset $A\subseteq M$, a subset $S_A\subseteq V(G)$ of minimum weight that dominates $A$.
\end{definition}

Next, we present our algorithms for {\sc Generalized Weighted $d$-Modulator Domination} and {\sc Weighted $d$-Decomposition Domination}.  For the {\sc Generalized Weighted $d$-Modulator Domination}  problem, we will use the following result. 

\begin{proposition}[\cite{cyganetal.book}, Implicit]\label{prop:setCoverExpAlgo}
The {\sc Generalized Weighted Set Cover} problem admits an $O(2^n\cdot m)$-time algorithm, where $n$ is the size of the universe and $m$ is the size of the set-family.
\end{proposition}

\begin{lemma}\label{lem:modulatorDom}
The {\sc Generalized Weighted $d$-Modulator Domination} problem admits an $O(2^{|M|}\cdot n)$-time algorithm.
\end{lemma}

\begin{proof}
We first describe the algorithm. Let $(G,w,M)$ be an instance of the {\sc Generalized Weighted $d$-Modulator Domination} problem. Then, we construct an instance $(U,{\cal F},w')$ of {\sc Generalized Weighted Set Cover} as follows:
\begin{itemize}
\item $U=M$.
\item ${\cal F}=\{N[v]\cap M: v\in V(G)\}$.
\item For every $F\in{\cal F}$, let $v_F$ be a vertex of minimum weight among the vertices $v\in V(G)$ that satisfy $F=N[v]\cap M$, and define $w'(F)=w(v_F)$.
\end{itemize}
We call the algorithm of Proposition \ref{prop:setCoverExpAlgo} with $(U,{\cal F},w')$ as input, and, for every $A\subseteq U$, let ${\cal S}_A$ be its output. Then, for every $A\subseteq M$, we return $S_A=\{v_F: F\in{\cal S}_A\}$.

Clearly, due to Proposition \ref{prop:setCoverExpAlgo}, the algorithm runs in $O(2^{|M|}\cdot n)$ time. For correctness, consider some $A\subseteq M$. Observe that, on the one hand, if $B\subseteq V(G)$ dominates $A$, then ${\cal B}=\{N[v]\cap A: v\in B\}\subseteq {\cal F}$ covers $A$ and $w(B)\geq w'(B)$. On the other hand, if ${\cal B}\subseteq {\cal F}$ covers $A$, then $B=\{v_F: F\in{\cal A}\}\subseteq V(G)$ dominates $A$, and $w'({\cal B})=w(B)$. This completes the proof.
\end{proof}

For the {\sc Weighted $d$-Decomposition Domination} problem, we will use Proposition \ref{prop:domSetTwAlgo} and the following result.

\begin{proposition}[\cite{DBLP:journals/siamcomp/Bodlaender96}]\label{prop:twComputation}
There exists an algorithm that, given a graph $G$, outputs a tree decomposition of $G$ of width $t=\mathsf{tw}(G)$ in $t^{O(t^3)}\cdot n$ time. 
\end{proposition}

\begin{lemma}\label{lem:decompDom}
The {\sc Weighted $d$-Decomposition Domination} problem admits an $O(2^{|M|}\cdot n)$-time algorithm.
\end{lemma}

\begin{proof}
We first describe the algorithm. Let $(G,w,M)$ be an instance of the {\sc Weighted $d$-Decomposition Domination} problem. Then, for every subset $L\subseteq M$, we construct an instance $I_L=(G_L,w_L,{\cal T}_L)$ of {\sc Weighted Dominating Set} parameterized by $\mathsf{tw}$ as follows:
\begin{itemize}
\item Let $V(G_L)=(V(G)\setminus M)\cup\{x\}$ for $x\notin V(G)$, and $E(G_L)=E(G-M)\cup\{\{x,v\}: v\in N_G(L)\setminus M\}$. That is, we construct $G_L$ from $G$ by removing the vertices in $M$ and the edges incident to them, and adding a new vertex $x$ adjacent to all of the vertices in $N_G(L)\setminus M$.
\item For every $v\in V(G_L)$, define $w_L(v)=w(v)$ if $v\in V(G)\setminus M$, and $w_L(v)=w(L)$ otherwise (for $v=x$).
\item Use the algorithm of Proposition \ref{prop:twComputation} with $G_L$ as input, and let ${\cal T}_L$ be its output.
\end{itemize}
Let ${\cal I}=\{I_L: L\subseteq M\}$. For every $I_L\in{\cal I}$, we call the algorithm of Proposition \ref{prop:domSetTwAlgo} with $I_L$ as input, let $S'_L$ be its output, and define $S_L$ as $S'_L$ if $x\notin S'_L$ and $S'_L\cup L$ otherwise. Let ${\cal S}=\{S_L: L\subseteq M\}$. Then, we return the set $S$ of minimum-weight with respect to $w$ among the sets in $\cal S$.

For the time complexity analysis, observe that $|{\cal I}|=2^{|M|}$. Moreover, observe that for every $L\subseteq M$, $\mathsf{tw}(G_L)\leq \mathsf{tw}(G-M)+1\leq d+1$; hence, each call to the algorithm of Proposition \ref{prop:twComputation} runs in $(d+1)^{O((d+1)^{d+1})}\cdot |V(G_L)|\leq O(n)$ time, and each call to the algorithm of Proposition \ref{prop:domSetTwAlgo} runs in $O(3^{d+1}\cdot |V(G_L)|)\leq O(n)$ time. Thus, the total running time of our algorithm is $O(2^{|M|}\cdot n)$.

Now, we turn to consider the correctness of the algorithm. To this end, consider some subset $L\subseteq M$. On the one hand, consider some subset $A\subseteq V(G)$ that satisfies $A\cap M=L$ and $A$ dominates $V(G)\setminus M$.
Let $A'=(A\setminus M)\cup\{x\}$. Then, $A\setminus M$ dominates $V(G_L)\setminus N_G(L)$ and $x$ dominates $N_G(L)$, hence $A'$ dominates $V(G_L)$, and our definition of $w_L$ directly implies that $w(A)=w_L(A')$.
On the other hand, consider some subset $A'\subset V(G_L)$ that dominates $V(G_L)$. Then, define $A$ as $A'$ if $x\notin A'$ and $A'\cup L$ otherwise. So, it is easy to see that $A$ dominates $V(G)\setminus M$ and $w_L(A')=w(A)$.

We conclude that, on the one hand, if $A\subseteq V(G)$ dominates $M$, then, for $L=A\cap M$, a minimum-weight dominating set of $G_L$ with respect to $w_L$ is of weight $w(A)$. So, the output dominating set cannot have weight larger than $w(A)$. On the other hand, for every $L\subseteq M$, the minimum weight of a dominating set of $G_L$ with respect to $w_L$ is  bounded from below by the minimum weight of a dominating set of $G$ with respect to $w$. So, obviously, the output dominating set cannot have weight larger than the minimum one. This completes the proof.
\end{proof}

Now, we are ready to conclude the correctness of Theorem \ref{thm:constTwAlgo}.

\begin{proof}[Proof of Theorem \ref{thm:constTwAlgo}]
We first describe the algorithm. Let $(G,w,M)$ be an instance of {\sc Weighted Dominating Set} parameterized by $\mathsf{tw}_d$. Then, we call the algorithms of Lemmas \ref{lem:modulatorDom} and \ref{lem:decompDom} with $(G,w,M)$ as input, and let $S_1$ and $S_2$ denote their outputs. We return $S=S_1\cup S_2$.

Clearly, due to Lemmas \ref{lem:modulatorDom} and \ref{lem:decompDom}, and since $|M|=\mathsf{tw}_d$, the algorithm runs in  $O(2^{\mathsf{tw}_d}\cdot n)$ time. For correctness, first note that due to the second item in Observation \ref{obs:mergeProblems}, the output set $S$ is a dominating set of $G$. Moreover, due to the first item in Observation \ref{obs:mergeProblems}, the optimum of $(G,w,M)$ as an instance of {\sc Weighted $d$-Modulator Domination} (or {\sc Weighted $d$-Decomposition Domination}) is bounded from above by the optimum of $(G,w,M)$ as an instance of {\sc Weighted Dominating Set}. Hence, both of $w(S_1),w(S_2)$ are bounded from above by the optimum of $(G,w,M)$ as an instance of {\sc Weighted Dominating Set}, which implies that $w(S)$ is bounded from above by twice the optimum of $(G,w,M)$ as an instance of {\sc Weighted Dominating Set}. This completes the proof.
\end{proof}

\section{Parameter: Vertex Cover Number}\label{sec:vc}
%\section{{\sc Dominating Set} Parameterized by Vertex Cover Number}\label{sec:vc}

In this section, we prove that in the case of $\mathsf{vc}=\mathsf{tw}_0$, we can attain an exact algorithm with the same running time as in Theorem \ref{thm:constTwAlgo}.

\begin{theorem}\label{thm:vcAlgo}
The {\sc Weighted Dominating Set} problem parameterized by $\mathsf{vc}$ admits a  $O(2^{\mathsf{vc}}\cdot n)$-time algorithm.
\end{theorem}

\begin{proof}
We suppose that the input also consists of a subset $M\subseteq V(G)$ that is a vertex cover of $G$ of size $\mathsf{vc}$, since such a subset can be easily computed in  $O(2^{vc}\cdot n)$ time~\cite{downey.fellows,cyganetal.book}. Now, we describe the algorithm. Let $(G,w,M)$ be an instance of {\sc Weighted Dominating Set} parametrized by $\mathsf{vc}$. We perform the following steps:
\begin{enumerate}
\item Call the algorithm of Lemma \ref{lem:modulatorDom} with $(G,w,M)$ as input of {\sc Weighted $0$-Modulator Domination}. Let $\{\widetilde{S}_A: A\subseteq M\}$ be its output.
\item For every $A\subseteq M$:
	\begin{enumerate}
	\item Let $\widehat{S}_{A}=A\cup (V(G)\setminus (N_G(A)\cup M))$. That is, $\widehat{S}_{A}$ is the union of $A$ and the set of vertices in the independent set $V(G)\setminus M$ that are not dominated by the vertices in $A$.
	\item Let $S_{A}=\widehat{S}_{A}\cup \widetilde{S}_{M\setminus N_G[\widehat{S}_A]}$. Notice that $M\setminus N_G[\widehat{S}_A]$ is the set of vertices in $M$ that are not dominated by the vertices in $\widehat{S}_{A}$. 
	\end{enumerate}
\item Return the set $S$ of minimum weight among the sets in $\{S_A: A\subseteq M\}$.
\end{enumerate}

Clearly, due to Lemma \ref{lem:modulatorDom}, the algorithm runs in  $O(2^{\mathsf{vc}}\cdot n)$ time. Moreover, it is clear that the output set $S$ is a dominating set of $G$. So, it remains to show that $S$ is of minimum weight among all dominating sets of $G$. To this end, let $S^\star$ be a dominating set of $G$ of minimum weight. Consider the iteration of the algorithm that corresponds to $A^\star=S^\star\cap U$. Notice that, since $S^\star$ dominates $V(G)\setminus M$ which is an independent set,  it must hold that $V(G)\setminus (N_G(A^\star)\cup M)\subseteq S^\star$. So, $\widehat{S}_{A^\star}\subseteq S^\star$. Further, since $S^\star$ dominates $M\setminus N_G[\widehat{S}_{A^\star}]$, we have that $S^\star\setminus\widehat{S}_{A^\star}$ dominates $M\setminus N_G[\widehat{S}_{A^\star}]$. By the correctness of the algorithm of Lemma \ref{lem:modulatorDom}, this implies that $w(\widetilde{S}_{M\setminus N_G[\widehat{S}_{A^\star}]})\leq w(S^\star\setminus \widehat{S}_{A^\star})$. Thus, we conclude that $w(S)\leq w(S_{A^\star})\leq w(S^\star)$.
\end{proof}

Additionally, we observe that the time complexity in Theorem \ref{thm:constTwAlgo} is tight.

\begin{observation}\label{thm:vcAlgoLower}
Under any of the SETH and the Set Cover Conjecture, the {\sc Dominating Set} problem parameterized by $\mathsf{vc}$ cannot be solved in  $O((2-\epsilon)^{\mathsf{vc}}\cdot n)$ time for any fixed $\epsilon>0$.
\end{observation}

\begin{proof}
First, we present a reduction from {\sc Hitting Set} to {\sc Dominating Set}. The input of {\sc Hitting Set} consists of a universe $U$ and a family ${\cal F}\subseteq 2^U$ of subsets of $U$, and the objective is to output a minimum-size subuniverse $U'\subseteq U$ such that for every set $F\in{\cal F}$, $U'\cap F\neq\emptyset$. We suppose that $|U|\leq|{\cal F}|$. It is known that under the SETH, for every fixed $\epsilon>0$, {\sc Hitting Set} cannot be solved in $O((2-\epsilon)^n)$ time where $n$ is the size of the universe~\cite{DBLP:journals/talg/CyganDLMNOPSW16}.
Given an instance $(U,{\cal F})$ of {\sc Hitting Set}, we construct an instance $G$ of {\sc Dominating Set} as follows: $V(G)=U\cup\{v_F: F\in{\cal F}\}\cup\{x,y\}$; $E(G)=\{\{u,v_F\}: u\in F\}\cup\{\{u,x\}:u\in U\}\cup\{\{y,x\}\}$. It is easy to see that the minimum size of a solution to $(U,{\cal F})$ equals the minimum size of a solution to $G$ minus $1$. Since $U\cup\{y\}$ is a vertex cover of $G$, the first part observation follows.

Second, we present a similar reduction from {\sc Set Cover} to {\sc Dominating Set}. For this purpose, let $(U,{\cal F})$ be an instance of {\sc Set Cover}. We suppose that $|U|\leq|{\cal F}|$. Then, we construct an instance $G$ of {\sc Dominating Set} as follows: $V(G)=U\cup\{v_F: F\in{\cal F}\}\cup\{x,y\}$; $E(G)=\{\{u,v_F\}: u\in F\}\cup\{\{x,v_F\}:F\in{\cal F}\}\cup\{\{x,y\}\}$. It is easy to see that the minimum size of a solution to $(U,{\cal F})$ equals the minimum size of a solution to $G$ minus $1$. Since $U\cup\{x\}$ is a vertex cover of $G$, the second part of the observation follows.
\end{proof}

\section{Parameter: Feedback Edge Set Number: FPT Algorithm}\label{sec:fes}

%\section{{\sc Dominating Set} Parameterized by Feedback Edge Set Number: FPT Algorithm}\label{sec:fes}

In this section, we first prove a combinatorial result (stated in Lemma \ref{lem:fesByTw2}). In particular, this result implies a parameterized algorithm where the basis of the exponent is smaller than $3$ (stated in Theorem \ref{thm:fesAlgo}). For our combinatorial result, we will use the following proposition.

\begin{proposition}[\cite{DBLP:journals/tcs/Bodlaender98}]\label{prop:cactusTw}
The treewidth of a cactus is at most $2$.
\end{proposition}

\begin{lemma}\label{lem:fesByTw2}
For any graph $G$, $\mathsf{tw}_2(G)\leq \frac{\mathsf{fes}(G)}{2}$.
Moreover, there exists an algorithm that, given a graph $G$, outputs a subset $M\subseteq V(G)$ such that $|M|\leq \frac{\mathsf{fes}(G)}{2}$ and $\mathsf{tw}(G-M)\leq 2$ in  $O(\mathsf{fes}(G)+n)$ time.
\end{lemma}

The idea behind the algorithm presented in the proof is quite simple (though, perhaps, if we did not demand it to run in  $O(\mathsf{fes}(G)+n)$ time, it could have been further simplified). Specifically, we scan a depth-first search (DFS) tree $T$ of $G$ from top to bottom. For each vertex that we remove (and insert into $M$), we aim to argue that at least two edges in $F=E(G)\setminus E(T)$ have become ``irrelevant'' ---that is, not part of any cycle. To identify which vertices to remove, we maintain a variable $e$, which stores an edge from $F$ whose ``top'' is above (or equal to) and whose ``bottom'' is below (or equal to) the vertex currently under consideration, and, most importantly, which is still ``relevant''. When no such edge exists, it stores $\mathsf{nil}$. In particular, we notice two situations where we can (and it suffices) to remove a vertex: first, when it is the top of two edges from $F$, and second, when it is the top of an edge from $F$ and $e$ is some other edge from $F$. We now proceed to present the formal description of the algorithm and its proof.

\begin{proof}[Proof of Lemma \ref{lem:fesByTw2}]
To describe the algorithm, let $G$ be a graph. Without loss of generality, we suppose that $G$ is connected, else we can consider each of its connected components separately.  We compute a DFS tree $T$ of $G$. Let $F=E(G)\setminus E(T)$, and note that $|F|=\mathsf{fes}(G)$. Given an edge $e\in F$, we refer to the {\em top} and {\em bottom} of $e$ as the endpoint of $e$ that is closer to the root of $T$ and the other endpoint of $e$, respectively. (Since $T$ is a depth-first search tree, the  terms top and bottom are uniquely defined.) 

Initialize $M=\emptyset$ and $e=\mathsf{nil}$. For every $v\in V(T)$ where $T$ is traversed in preorder, we perform the following computation:
\begin{enumerate}
\item If $v$ is the bottom of $e$ (in this case, $e\neq\mathsf{nil}$), update $e=\mathsf{nil}$.
\item If $v$ is not the top of any edge in $F$, we proceed to the next iteration.
\item If either $v$ is the top of at least two edges in $F$ or  $e\neq\mathsf{nil}$, then:
	\begin{enumerate}
	\item Insert $v$ into $M$.
	\item Update $e=\mathsf{nil}$.
	\item Proceed to the next iteration.
	\end{enumerate}
\item Update $e$ to be the edge in $F$ whose top is $v$, and prioritize the preorder traversal to first visit the vertices on the subpath of $T$ from $v$ to the bottom of $e$. (In case a previous prioritization exists, override it.)
\end{enumerate}
At the end, we return the set $M$.

Clearly, the algorithm runs in  $O(n+m)=O(\mathsf{fes}(G)+n)$ time.

We now turn to consider the correctness of the algorithm. Towards that, we define the following terminology. Given an edge $e\in F$, let the {\em span} of $e$ be the subpath of $T$ from the top to bottom of $e$, and let the {\em truncated span of $e$} be the subpath that results from the removal of the bottom of $e$ from the span of $e$.  We say that two distinct edges $e,e'\in F$ have a {\em conflict} if the top of one of them belongs to truncated span of the other.  
Given an edge $e\in F$ and a subset $M\subseteq V(G)$, we say that $e$ is {\em active} in $M$ if $G-M$ contains a cycle that traverses $e$. Observe that all of the edges in $F$ are active in $\emptyset$.

Towards the proof of our main inductive claim, we present the following claim.

\begin{claim}\label{claim:sameCycle}
Let $C$ be a cycle in $G$, $e\in E(C)\cap F$, and suppose that $C$ is not the cycle formed by $e$ and its span. Then, $C$ contains an edge $e'\in F\setminus\{e\}$ that has a conflict with $e$, and whose top is either the top of $e$ or an ancestor of it.
\end{claim}

\begin{proof}
Let $t$ and $b$ be the top and bottom of $e$, respectively. Targeting a contradiction, we assume that $C$ does not contain an edge $e'\in F\setminus\{e\}$ that has a conflict with $e$, and whose top is either $t$ or an ancestor of $t$.  Let $P$ denote the subpath of $C$ between $t$ and $b$ that does not contain $e$. Due to our assumption, this path cannot contain an edge between $t$ or an ancestor of $t$ and a descendant of $t$, with the exception of the edge between $t$ and its children in $T$, because such an edge must belong to $F$ and have a conflict with $e$. Due to this, and because $T$ is a depth-first tree, $P$ cannot contain an edge between a vertex that is not a descendant of $t$ and a descendant of $t$, with the exception of the edge between $t$ and its child that  belongs to the span of $e$, which we denote by $c$. So far, we conclude that $P$ does not contain any ancestor of $t$ and that it contains the edge $\{t,c\}$. However, again, because $T$ is a depth-first tree, $P$ also cannot contain an edge between a vertex that is a descendant of a vertex, say, $x$, that belongs to the span of $e$ and a vertex that is neither $x$ nor an ancestor of $x$. In turn, this implies that $P$ is equal to the span of $e$, which is a contradiction to the supposition of the claim that $C$ is not the cycle formed by $e$ and its span. 
\end{proof}

Now, we are ready to present our main inductive argument.

\begin{claim}\label{claim:inductive}
Consider an iteration of the preorder traversal. Let $M'$ be the set $M$ at the end of this iteration. Let $e'$ denote the value of $e$ at the end of this iteration. Let $v$ be the vertex traversed in this iteration. Then:
\begin{enumerate}
\item\label{item1} The set $M'$ does not contain any descendant of $v$.
\item\label{item2} There do not exist two  edges in $F$ that are active in $M'$, have a conflict and the top of each one of them is either $v$ or an ancestor of $v$ in $T$.
\item\label{item3}  If $e'=\mathsf{nil}$, then there does not exist an edge in $F$ that is active in $M'$ and such that $v$ belongs to the truncated span of that edge.
\item\label{item4} If $e'\neq\mathsf{nil}$, then: (i) $v$ belongs to the truncated span of $e'$; (ii) there does not exist an edge in $F$ other than $e'$ that is active in $M'$ and such that $v$ belongs to the truncated span of that edge; (iii) $M'$ does not contain any vertex from the span of $e'$. (In particular due to item \ref{item1} and (iii), $e'$ is active, and this is witnessed by the cycle formed by $e'$ and its span.)
\end{enumerate}
\end{claim}

\begin{proof}
We use induction on the preorder traversal. Consider the first iteration, where $v$ is the root of $T$ and, hence, the only edges in $F$ such that $v$ belongs to their span are those that have $v$ as their top. Then, all of the items in the claim directly follow from the pseudocode.

Now, consider an iteration that is not the first, and suppose that the claim is correct up to this iteration. By the inductive hypothesis (item \ref{item1}) and the pseudocode, it should be clear that item \ref{item1} of the claim holds.  Let $M''$ and $e''$ denote the values of $M$ and $e$ at the beginning of the iteration.  Let $u$ be the parent of $v$ in $T$. By the inductive hypothesis (item \ref{item2}), there do not exist two  edges in $F$ that are active in $M''$, have a conflict and the top of each one of them is an ancestor of $u$ in $T$. Yet, to prove item \ref{item2} of the claim, we still need to argue that there do not exist two  edges in $F$ that are active in $M'$, have a conflict, the top of one of them is $v$, and the top of the other is either $v$ or an ancestor of $v$. Note that if there exist two such edges, then $v$ belongs to the truncated span of both of these edges. We consider the two following cases.
\begin{itemize}
\item First, suppose that $e''=\mathsf{nil}$. Then, by the inductive hypothesis (item \ref{item3}), there does not exist an edge in $F$ that is active in $M''$ and such that $u$ belongs to the truncated span of that edge. So, the only edges that are active in $M''$ and such that $v$ belongs to their span are those that have $v$ as their top. If $v$ is not the top of any edge in $F$, then $e'=\mathsf{nil}, M'=M''$, and items \ref{item2} and \ref{item3} of the claim follow. If $v$ is the top of at least two edges in $F$, then $e'=\mathsf{nil}, M'=M''\cup\{v\}$ (so, these edges are non-active in $M'$), and items \ref{item2} and \ref{item3} of the claim follow. If $v$ is the top of exactly one edge in $F$, then this edge is $e'$ (and $M'=M''$), and, hence, items \ref{item2} and \ref{item4} of the claim follow.

\item Second, suppose that $e''\neq\mathsf{nil}$. Then, by the inductive hypothesis (item \ref{item4}), $u$ belongs to the truncated span of $e''$, there does not exist an edge in $F$ other than $e''$ that is active in $M''$ and such that $u$ belongs to the truncated span of that edge, and $M''$ does not contain any vertex from the span of $e''$. We further consider the three following sub-cases:
	\begin{enumerate}
	\item First, suppose that $v$ is the bottom of $e''$. This implies that the only edges that are active in $M''$ and such that $v$ belongs to their span are those that have $v$ as their top. Then, $e$ is updated to be $\mathsf{nil}$ in the first step of the iteration, and the proof proceeds as in the first case.
	\item Second, suppose that $v$ is neither the bottom of $e''$ nor the top of any edge in $F$. This implies that $v$ belongs to the truncated span of $e''$, and that there does not exist an edge in $F$ other than $e''$ that is active in $M''$ and such that $v$ belongs to the span of that edge.  As $e'=e''$ and $M'=M''$, items \ref{item2} and \ref{item4} of the claim follow.
	\item Third, suppose that $v$ is not the bottom of $e''$, and that $v$ is the top of at least one edge in $F$. Then, $e'=\mathsf{nil}$ and $M'=M''\cup\{v\}$. Hence, there does not exist an edge in $F$ that is active in $M'$ and has $v$ as its top. Hence, item \ref{item2} of the claim follows, and to complete the proof of item \ref{item4} of the claim, it suffices to show that $e''$ is non-active in $M'$. 
	
Targeting a contradiction, suppose that $e''$ is active in $M'$, and let $t$ and $b$ denote its top and bottom, respectively. Then, there exists a cycle $C$ in $G-M'$ that contains $e''$. In particular, there exists a path $P$ in $G-M'$ between $t$ and $b$ that does not contain $e''$. Due to Claim \ref{claim:sameCycle}, if $P$ is not equal to the span of $e''$, then $C$ contains an edge $\widehat{e}\in F\setminus\{e''\}$ that has a conflict with $e''$ and whose top is either $t$ or an ancestor of $t$, and because this edge belongs to $C$ (which exists in $G-M'$), it must be active in $M'$; however, this is a contradiction to item \ref{item2} of the claim. Thus, $P$ is equal to the span of $e''$, which is a contradiction, since $v$ belongs to this span as well as to $M'$. So, $e''$ is non-active in $M'$.
%	Targeting a contradiction, suppose that $e''$ is active in $M'$, and let $t$ and $b$ denote its top and bottom, respectively. Then, there exists a cycle $C$ in $G-M'\cup\{v\}$ that contains $e''$. In particular, there exists a path $P$ in $G-M'\cup\{v\}$ between $t$ and $b$ that does not contain $e''$. Due to item \ref{item2}, this path cannot contain an edge between $t$ or an ancestor of $t$ and a descendant of $t$, with the exception of the edge between $t$ and its children in $T$, because such an edge must belong to $F$, be active in $M'$ (being part of $C$) and have a conflict with $e''$. Due to this, and because $T$ is a depth-first tree, $P$ cannot contain an edge between a vertex that is not a descendant of $t$ and a descendant of $t$, with the exception of the edge between $t$ and its child that  belongs to the span of $e''$, which we denote by $c$. So far, we conclude that $P$ does not contain any ancestor of $t$ and that it contains the edge $\{t,c\}$. However, again, because $T$ is a depth-first tree, $P$ also cannot contain an edge between a vertex that is a descendant of a vertex, say, $x$, that belongs to the span of $e''$ and a vertex that is neither $x$ not an ancestor of $x$. In turn, this implies that $P$ is equal to the span of $e''$, which is a contradiction, since $v$ belongs to this span as well as to $M'$. So, $e''$ is non-active in $M'$.
	\end{enumerate}
\end{itemize}
This completes the proof.
\end{proof}

We proceed to prove the following claim, which will imply the desired bound the size of $M$.

\begin{claim}\label{claim:deactivate}
Consider an iteration of the preorder traversal. Let $M'$ be the set $M$ at the end of this iteration. Suppose that in this iteration, the vertex $v$ was inserted into $M'$. Then, there exist two distinct edges in $F$ that are active in $M'\setminus\{v\}$ but are non-active in $M'$.
\end{claim}

\begin{proof}
Let $e''$ denote the value of $e$ at the start of this iteration. Then, one of the two following cases holds.

\subparagraph*{Case I.} Suppose that $v$ is the top of at least two edges in $F$, say, $e^v_1$ and $e^v_2$. Then, due to item \ref{item1} of Claim \ref{claim:inductive}, both of these edges are active in $M'\setminus\{v\}$ (witnesses by the cycles formed by these edges and their spans). However, both of these edges clearly become non-active in $M'$.

\subparagraph*{Case II.} Suppose that $e''\neq\mathsf{nil}$ and $v$ is the top of exactly one edge in $F$,  denoted by $e^v$. Note that $e''\neq e^v$, since the value of $e$ is updated when its top is traversed (and $v$ is only being traversed in the current iteration, after $e$ already holds $e''$). As in Case I, $e^v$ is active in $M'\setminus\{v\}$ but becomes non-active in $M'$. By item \ref{item4} of Claim \ref{claim:inductive} with respect to the previous iteration, $e''$ is active in $M'\setminus\{v\}$, and by the same item with respect to the current iteration,  $e''$ is non-active in $M'$.

\smallskip\noindent
In either case, we conclude that the claim holds.
\end{proof}

In particular, from Claim \ref{claim:deactivate} we conclude that $|M|\leq \frac{|F|}{2}=\frac{\mathsf{fes}(G)}{2}$. (For every vertex inserted into $M$, at least two edges in $F$ that are active at that moment become non-active, and they never become active again later).

In order to bound the treewidth of $G-M$, we turn to prove several additional claims.

\begin{claim}\label{claim:fesActiveBound}
Let $X\subseteq V(G)$. Then, $\{e\in F: e$ is active in $X\}$ is a feedback edge set of $G-X$.
\end{claim}

\begin{proof}
The claims directly follows from the definition of active edges, and because $F$ is a feedback edge set of $G$.
\end{proof}

\begin{claim}\label{claim:cyclesCommon}
Let $X\subseteq V(G)$. Let $C,C'$ be two distinct cycles in $G-X$ that have at least two vertices in common. Then, there exist two distinct edges $e,e'\in F$ that are active in $X$ and have a conflict.
\end{claim}

\begin{proof}
By Claim \ref{claim:sameCycle} and since $C,C'$ belong to $G-X$, we can assume that $C$ and $C'$ are the cycles that consist of some edges $e,e'\in F$ and their spans, respectively, else the proof is complete. Since $C$ and $C'$ have at least two vertices in common, the intersection of the spans of $e$ and $e'$ must be of size at least $2$. However, this implies that the top of one of them must belong to the truncated span of the other, and hence they have a conflict.
\end{proof}

\begin{claim}\label{claim:noConflicts}
There do not exist two distinct edges $e,e'\in F$ that are active in $M$ and have a conflict.
\end{claim}

\begin{proof}
The claim directly follows from item \ref{item2} of Claim \ref{claim:inductive} by considering the iterations in which the leaves of $T$ were traversed.
\end{proof}

From Claims \ref{claim:cyclesCommon} and \ref{claim:noConflicts}, we derive that $G-M$ is a cactus graph. So, by Proposition \ref{prop:cactusTw}, we conclude that its treewidth is at most $2$. This completes the proof.
\end{proof}

\begin{theorem}\label{thm:fesAlgo}
The {\sc Weighted Dominating Set} problem parameterized by $\mathsf{fes}$ admits an $O(3^{\frac{\mathsf{fes}}{2}}\cdot n)$-time algorithm.
\end{theorem}

\begin{proof}
To describe the algorithm, let $(G,w)$ be an instance of {\sc Weighted Dominating Set} . Then, we call the algorithm of Lemma \ref{lem:fesByTw2}, and let $M$ be its output. So, $|M|\leq \frac{\mathsf{fes}(G)}{2}$ and $\mathsf{tw}(G-M)\leq 2$. Afterwards, we call the algorithm of Proposition \ref{prop:twComputation} with $G-M$ as input, and let ${\cal T}'$ be its output. So, ${\cal T}'$  is a tree decomposition of width at most $2$ of $G-M$. We insert $M$ into each of the bags of ${\cal T}'$ to attain a tree decomposition ${\cal T}$ of $G$ of width at most $|M|+2\leq \frac{\mathsf{fes}(G)}{2}+2$. Lastly, we call the algorithm of Proposition \ref{prop:domSetTwAlgo} with $(G,w,{\cal T})$ as input, and return its result.

Clearly, correctness is immediate. As for the time complexity, observe that the calls to the algorithms of Lemma \ref{lem:fesByTw2} and Propositions \ref{prop:twComputation} and \ref{prop:domSetTwAlgo} run in $O(\mathsf{fes}(G)+n)$, $2^{O(2^3)}\cdot n=O(n)$ and $O(3^{\frac{\mathsf{fes}(G)}{2}+2}\cdot n)\leq O(3^{\frac{\mathsf{fes}(G)}{2}}\cdot n)$ times, respectively. Thus, the total running time of our algorithm is $O(3^{\frac{\mathsf{fes}(G)}{2}}\cdot n)$.
\end{proof}

\section{Parameter: Feedback Edge Set Number: Compression Algorithm}\label{sec:compression_}

In order to state our compression result, we introduce a generalization of {\sc Dominating Set}.

\begin{definition} [$\{G,W\}$-Dominating Set]
Let $G$ be a graph and $W\subseteq V(G)$. We call a set $S\subseteq V(G)$ a {\em $\{G,W\}$-dominating set} if $S$ dominates the vertices in $V(G)-W$ and any number of vertices in $W$. When $G$ is clear from context, we use the abbreviation $W$-dominating set.
\end{definition}

\begin{definition}
    [Relaxed Dominating Set ({\sc RDS}) instance]
We denote by $\{G,W\}$ a pair of a graph $G$ and a set of vertices $W\subseteq V(G)$, and call it an {\em {\sc RDS} instance}. 
Note that $\{G,\emptyset\}$ is an instance of the standard dominating set problem. 
\end{definition}

%\begin{definition} [{\sc RDS: Relaxed Dominating Set Problem}] ~\\
% {\em Input:} A pair $\{W,G\}$ of a graph $G$ and a set $W\subset V(G)$. \\
% {\em Output:} A minimum-size $\overline{W}$-dominating set 
%\end{definition}

\begin{definition} [{\sc RDS} Reduction]
Let $\{G ,W\}, \{G',W'\}$ be two {\sc RDS} instances with $V(G') \subseteq V(G)$. 
We say that $\{G ,W\}$ {\em reduces} to $\{G',W'\}$ if we can compute a set $S\subseteq V(G) \setminus V(G')$ such that the union of $S$ and a minimum $\{G',W'\}$-dominating set, is a minimum $\{G,W\}$-dominating set. 
\end{definition}

\begin{theorem} \label{thm:compression}
 Let $G$ be a graph that has FES number equal to $k$. Then $\{G,\emptyset\}$ can be reduced in linear time to an RDS instance $\{G', W'\}$  where $G'$ has size $O(k)$. \footnote{More specifically, $G'$ has at most $2k$ vertices of degree more than 2. It also has at most $27k$ edges and at most $26k$ vertices, which is subject to improvement.} 
\end{theorem}

\subsection{A Sketch of the Proof}

We now give a sketch of the proof.
%that is contained in the Appendix.
At a high level, the proof is based on what we call a {\em Cactus-Kernel decomposition} of the graph $G$, described in Claim~\ref{claim:decomposition_} that implicitly used in other works~\cite{gupta04cuts}; for completeness we prove it in Section~\ref{sec:cactus-kernel}.~\footnote{The decomposition itself is based on a well-known greedy vertex elimination algorithm that likely has its origins in algorithms for solving linear systems~(e.g. see~\cite{koutis_linear_2007}), but also has been used to obtain other algorithmic results~\cite{gupta04cuts}.} 

\begin{claim} [Cactus-Kernel Decomposition] \label{claim:decomposition_}
 Let $G$ be a connected graph such that $|E| = |V|-1+k$, where $E=E(G)$ and $V=V(G)$.  Then $E$ can decomposed into two disjoint edge sets $E_{k}, E_{c}$ spanning vertices $V_k\subseteq V$ and $V_c \subseteq V$ respectively, such that:
 \begin{itemize}
     \item $G_{c} = (V_{c},E_{c})$ is a forest of cactus graphs.
     \item $G_{k} = (V_k, E_k)$, the `kernel', has a set $\tilde{V}_{k}$ of at most $2k$ vertices of degree greater than 2.
     \item $G_k$ consists of at most $3k$ edge-disjoint paths, whose endpoints are vertices of $\tilde{V}_{k}$.  
     \item Each cactus in $G_c$ contains a single vertex from $V_k$. We call it the {\em root} of that cactus.
\end{itemize}     
\end{claim}

The Cactus-Kernel decomposition can be computed in linear time. The intuition with the decomposition is that it identifies in $G$ multiple induced paths, i.e. paths spanned by their vertices, and edge-disjoint cycles that enable essentially greedy algorithms for their optimal domination in $G$; a number of related Lemmas are provided in Section~\ref{sec:path-cycle-domination}. The algorithm consists of the following steps:
\medskip

\noindent \textbf{(a)}: Computing for each cactus $H$, a minimum set $S_H$ that dominates $H$ with the possible exception of its root.  $S_H$ has the property that there exists a minimum dominating set of $G$ that contains it.  The procedure for computing $S_H$ exploits the tree structure underlying the cactus; it is described in Section~\ref{sec:dangling}. Step (a) reduces the $\{G,\emptyset\}$ to a smaller instance $\{G_k', W\}$.  \\ 
\noindent \textbf{(b)}: Graph $G_k'$ has potentially long induced paths, that are replaced by fixed-length paths using a procedure described in Section~\ref{sec:path-reduction}. This reduces $\{G_k',W\}$, to $\{G',W'\}$, where $G'$ is as specified
in Theorem~\ref{thm:compression}.

\subsection{Cactus-Kernel Decomposition} \label{sec:cactus-kernel}

\subsubsection{Definitions and Claims}
Here we introduce definitions and lemmas related to the cactus-kernel decomposition claimed in Section~\ref{sec:compression_}. In this section we will be using the notation $G=(V,E)$ for a graph $G$ with $V=V(G)$ and $E=E(G)$.

\begin{definition} [Tree Structure of Cactus] \label{def:tree_structure}
Let $H$ be a cactus graph with a root vertex~$u$. Besides non-trivial cycles, we also consider as a cycle in $H$ each single vertex which does not belong to any non-trivial cycle. We define a rooted tree $T_c = (V_c,E_c)$ as follows: 
\smallskip

\noindent (a) The vertices in $V_c$ are in 1-1 correspondence to cycles in $G$. For $v\in V_c$ let $C(v)$ be the set of vertices in the cycle corresponding to $v$.  We pick an arbitrary vertex $u_c \in V_c$ such that $u \in C(u_c)$, and we designate $u_c$ as the root of $T_c$.
\\\noindent (b) If $u,w \in V_c$ then $(u,w) \in E_c$ if $C(u)$ and $C(w)$ are joined by an edge or if they share a vertex. 

%\hly{The unique vertex $u_c \in V_c$ such that $u \in C(u_c)$ is designated as the root of $T_c$}. We designate an arbitrary vertex $u_c \in V_c$ such that $u \in C(u_c)$ as the root of $T_c$.

\smallskip
\noindent For each $v \in V_c$ such that $C(v)$ contains $u$, we call $u$ the {\bf \em articulation vertex} of $C(v)$. For any other $v$, the cycle $C(v)$ contains a unique vertex $w$ that belongs to, or is connected via a single edge to $C(u)$, where $u$ is the parent of $v$ in $T_c$. We call $w$ the  articulation vertex of $C(v)$, and we respectively say that $C(v)$ is \textbf{vertex-attached} or \textbf{edge-attached} to $C(u)$. 
\end{definition}

\begin{claim} [Cactus-Kernel Decomposition] \label{claim:decomposition}
 Let $G = (V,E)$ be a connected graph such that $|E| = |V|-1+k$.  Then $E$ can decomposed into two disjoint edge sets $E_{k}, E_{c}$ spanning vertices $V_k\subseteq V$ and $V_c \subseteq V$ respectively, such that:
 \begin{itemize}
     \item $G_{c} = (V_{c},E_{c})$ is a forest of cactus graphs.
     \item $G_{k} = (V_k, E_k)$, the `kernel', has at most $2k$ vertices of degree greater than 2. We denote by $\tilde{V}_{k}$ these vertices.
     \item $G_k$ consists of at most $3k$ edge-disjoint paths, whose endpoints are vertices of $\tilde{V}_{k}$.  
     \item Each cactus in $G_c$ contains a single vertex from $V_k$, and we call this vertex the {\em root} of that cactus.
\end{itemize}     
\end{claim}

\begin{claim} \label{claim:tree_structure}
 The Cactus-Kernel decomposition of a graph can be computed in linear time. The same applies for the tree structure of a cactus graph.
\end{claim}

The rest of the section proves Claims~\ref{claim:decomposition} and~\ref{claim:tree_structure}. The proof is based on an greedy vertex elimination algorithm that is well known in the context of algorithms for solving linear systems, but it is properly adapted for our context. 

\subsubsection{The Elimination Algorithm and Its Properties} \label{sec:elim}

We begin with the elimination process, presented as Algorithm~\ref{alg:elimination}, that progressively eliminates nodes and vertices.

\begin{algorithm} 
\SetAlgoLined
 \textbf{Input:} Graph $G=(V,E)$\;
 $H=G$\; 
 %$S(v)=\{~\}$ for all $v\in V$\;
 %$P(e)=\{e\}$ and $S(e) = \{~\}$ for all $e\in E$\;
 
 \% {\small $P(e)$ is an ordered set of edges}\;
 \% {\small $C(v)$ is a collection of sets of vertices}\;
 %\% {\small $A$ is a set of vertices} \;
 \% {\small $B$ is set of edges}\;
 $P(e) = e$ for all $e\in E$\;
 ${\cal C}(v) = \{\}$ for all $v\in V$\;
 \While{H contains a vertex with degree at most 2}{
  let $v$ be a vertex with $deg(v)<3$ in $H$\;
  \% {\small \textbf{elimination type-A}}\;
  \If{deg(v)=1 and w is the neighbor of v}{
   eliminate node $v$ and the adjacent edge $e$\;
   add $(v,w)$ to $B$\;
   if ${\cal C}(v)=\emptyset$ let ${\cal C}(v)=\{\{v\}\}$}
   %$S(w) = S(w) \cup S(e) \cup S(v) \cup v$\;

  \% {\small \textbf{elimination type-B}}\; 
  \If{deg(v)=2 and v has two parallel edges to w}{
   eliminate node $v$ and its two adjacent edges $e_1,e_2$\;
   add $\{P(e_1) \cup P(e_2)\}$ to ${\cal C}(w)$\;}
   %let $S(e) = S(e_1) \cup S(e_2) \cup S(v) \cup v$\; 

  \% {\small \textbf{elimination type-C}}\; 
  \If{deg(v)=2 and u,w are the neighbors of v}{
   eliminate vertex $v$ and edges $e_1=(u,v)$ and $e_2=(v,w)$\;
   add an edge $e = (u,w)$\;
   %let $P(e) = [P(e_1), P(e_2)]$ and $S(e) = S(e_1) \cup S(e_2) \cup S(v) \cup v$\; 
   let $P(e) = [P(e_1), P(e_2)]$\;
   }
  \textbf{return} $H$, $P(e): \forall e \in H$, $C(v) \forall v \in V$
 }
 \caption{The elimination algorithm} \label{alg:elimination}
\end{algorithm}

 The following Lemmas analyze properties of the algorithm.

\begin{lemma} \label{lem:paths}
  Algorithm~\ref{alg:elimination} assigns to each edge $e$ an ordered set of edges $P(e)$ in $G$.  \\ (a) At any step of the algorithm, the set of edges $P(e), e\in V$ are mutually disjoint and their union contains a subset of the edges that have been eliminated up to that step.  \\ (b) For each $e$, $P(e)$ constitutes a simple path between the two endpoints of $e$ in $G$. 
\end{lemma}
\begin{proof}
  Observe that the claim is true at initialization, and all steps of the algorithm perform only unions of such sets.
  Hence (a) follows by a simple inductive argument. Similarly, in order to prove (b), if we inductively assume that 
  $P(e_1)$ and $P(e_2)$ respectively make paths between the endpoints of $e_1$ and $e_2$ respectively $G$, then $P(e)$ is the path we get by joining $P(e_1)$ and $P(e_2)$ at the eliminated vertex $v$.
\end{proof} 

\begin{lemma} (Eliminated Bridge) \label{lem:bridge}
  Let $G  = (V,E)$. Suppose $v \in V$ is eliminated via a Type-A elimination and $(v,w)$ is its adjacent edge. Then $(v,w)$ is a bridge in $G$. 
\end{lemma}
\begin{proof}
  For the sake of contradiction, suppose $(v,w)$ is on a cycle $c$ in $G$. 
  If $|c|=2$ then $c$ must consist of two parallel edges between $v$ and $w$. These edges cannot be eliminated before one of their endpoints is eliminated, which means that $v$ must have degree 2 just before its elimination, a contradiction. Suppose now we have $|c|>2$. Since $v$ has degree $2$ in $c$ and degree $1$ before its elimination, it follows that at least one more vertex in $c$ must be eliminated. Let $w$ be the first such vertex eliminated by the algorithm. Notice that before the elimination of $w$, $c$ is intact (by definition of $w$) and we must have $deg(w)=2$. Therefore, after the elimination of $w$ we get a new cycle $c'$ with $|c'|=|c|-1$. Inductively we get that at some point of the elimination process the cycle containing $(v,w)$ must have length 2, which as noted above contradicts the assumption of the Lemma. 
\end{proof}

\begin{lemma} [Eliminated Cycle] \label{lem:cycle}
  In a type-B elimination $P(e_1)\cup P(e_2)$ form a cycle in $G$. 
\end{lemma}
\begin{proof}
  The lemma follows directly from Lemma~\ref{lem:paths}
\end{proof}

\begin{lemma} (Eliminated Cycles are Edge-Disjoint) \label{lem:disjointCycle}
  Let $G  = (V,E)$. Let $c_1$ and $c_2$ be two cycles in $G$ that both have length at least 2 and share at least one edge. Then 
  the algorithm will leave uneliminated at least two vertices from each cycle. It follows that if all but one vertices of a cycle $c$ are eliminated, then its edges are not included in any other cycle. 
\end{lemma}
\begin{proof}
  For the sake of contradiction assume that all but one vertices of $c_1$ can be eliminated, and similarly for $c_2$.
  Suppose that one of the two cycles has length at least 3. Then, by assumption, we can find a vertex $v$ that is in at least one of the two cycles and gets eliminated by the algorithm. Before that elimination step, $c_1$ and $c_2$ are intact (by definition of $v$), and $v$ must have degree 2. Hence its elimination will still leave two cycles $c_1^{(1)}$ and $c_2^{(1)}$, such that $c_1^{(1)} \subseteq c_1$ and $c_2^{(2)} \subseteq c_2$, with at least one of the containments being proper; importantly $c_1^{(1)}$ and $c_2{(1)}$ will still share an edge. By applying the reasoning inductively, we get a sequence of containments $c_1^{(t)}\subseteq \ldots \subseteq c_1$ and
  $c_2^{(t)}\subseteq \ldots \subseteq c_2$, where both $c_1^{(t)}$ and $c_2^{(t)}$ have length 2 and share an edge. That can be the case only both $c_1^{(t)}$ and $c_2^{(t)}$ consist of two parallel edges. Since they also share one edge, it must that we have three parallel edges between two vertices $(u,v)$. These three parallel edges block the elimination of both of $u$ and $v$ for the rest of the process. Hence, the two vertices $u$ and $v$ (that belong to both $c_1$ and $c_2$) will remain uneliminated, and we get a contradiction.  
\end{proof}

The following is a well-known lemma. Here, we prove it for completeness.
\begin{lemma}  \label{lem:kernel} (Kernel Graph)
  Suppose a graph $G$ has n-1+k edges. Then
  the output $H$ of Algorithm~\ref{alg:elimination} has at most $3k$ edges and at most $2k$ vertices.
\end{lemma}
\begin{proof}
Suppose the elimination process takes $t$ iterations. Then $H$  has $n-1-t$ vertices and at most $n-1+k-t$ edges. The average degree of $H$ is then equal to $2(n-1+k-t)/(n-1-t)$.  The average degree of $H$ must be at least 3, otherwise the elimination process would not have stopped. We thus have
$$
      2(n-1+k-t) \geq  3(n-1-t) \Rightarrow t > n -1 - 2k.
$$
Hence the number of edges in $H$ is at most $3k$. Also, the number of vertices is at most $2k$. 
\end{proof}

Finally, we state the running time of the algorithm. 
\begin{lemma} \label{lem:elimrt}
  The running time of Algorithm~\ref{alg:elimination} is linear in the number of edges in $G$. 
\end{lemma}
\begin{proof} 
  Outside maintaining the sets $P(e), C(W), B$, the elimination process is known to be implementable in linear time, or even in parallel and with linear work~\cite{koutis_linear_2007}. Given that the sets $P(e)$ constitute paths (by Lemma~\ref{lem:paths}), then we can implement them as simple doubly-linked lists with $O(1)$ update time for each elimination step. 
 \end{proof}
 
\subsubsection{Completing the Proof of Claims~\ref{claim:decomposition}~and~\ref{claim:tree_structure}}

\begin{proof}
Let $H$ be the output of Algorithm~\ref{alg:elimination} on $G=(V,e)$. Each edge $e$ is associated with a path $P(e)$ between its endpoints in $G$, and $P(e)$ is disjoint from other such sets, by Lemma~\ref{lem:paths}. Replacing $P(e)$ with $e$ gives the graph $G_k=(V_k,E_k)$, where $\tilde{V}_k$ is the vertex set of $H$. The claims on the size of $G_k$ follow directly from lemma~\ref{lem:kernel}. 

Consider now all other edges in $E'=E-E_k$. A well known (and easy to show) property of the elimination algorithm is that the order of elimination does not affect the output of the algorithm. Thus, applying the elimination algorithm on the subgraph $G'$ of $G$ induced by the vertices of $E'$ would completely eliminate it. By Lemma~\ref{lem:disjointCycle}, all cycles in $G'$ are edge-disjoint, and thus $G'$ must be a forest of cactus graphs. 

Let's now focus on one cactus in the forest. By construction and Lemma~\ref{lem:disjointCycle}, each collection ${\cal C}(v)$ contains a set of edge-disjoint cycles, that all intersect at vertex $v$, and each cycle in the cactus appears only in a single ${\cal C}(w)$. It is also the case that each vertex $w$ that is not part of any cycle also has a non-empty collection ${\cal C}(w)$, due to the corresponding step for type-A elimination. Thus, non-empty collections ${\cal C}(v)$ are in 1-1 correspondence with the vertices $V_c$ tree structure $T_c = (V_c,E_c)$ of the cactus (Definition~\ref{def:tree_structure}). The edges of $T_c$ can be computed as follows: we first use all the bridges in set $B$. Then for each non-empty ${\cal C}(v)$ we scan all vertices in the union of the cycles in ${\cal C}(v)$, and find all vertices $w$, such that ${\cal C}(w)$ is non-empty. We then add to $T_c$ edges between the corresponding vertices in $V_c$. 

The running time of Algorithm~\ref{alg:elimination} is linear. It is also straightforward to see that the processing of its output for building $G_k$ and the cactus graphs, also takes linear time. 
\end{proof}

\subsection{Compression Algorithm}\label{sec:compression}

%\subsection{Relaxed Dominating Set}

%We introduce a generalization of dominating set. 

%\begin{definition} [{\sc RDS:  Relaxed Dominating Set Problem}] ~\\
%   {\em Input:} A graph $G$ and $W\subset V(G)$ \\
%   {\em Output:} A minimum-size set $S\subseteq V(G)$ that dominates all vertices in $V(G)-W$ and any number of vertices in $W$.
%\end{definition}

%\begin{definition} [{\sc RDS} Reduction]
%Let $\{G ,W\}, \{G',W'\}$ be two instances of {\sc RDS}, such that $V(G') \subseteq V(G)$. We say that $\{G ,W\}$ reduces to $\{G',W'\}$ if a minimum $W'$-dominating set for $G'$ is contained in a minimum $W$-dominating set for $G$. 
%\end{definition}

\medskip

\noindent \textbf{\large Notation:} In the rest of the section we will use the notation $G=(V,E)$ where $V=V(G)$ and $E=E(G).$

\subsubsection{Path \& Cycle Relaxed Domination} \label{sec:path-cycle-domination}

\begin{lemma} [Leaf Reduction Operation] \label{lem:leaf}
 Let $\{G,W\}$ be an instance of the {\sc RDS} problem. Let $v\in V$ a leaf vertex in $G$. Then (a) If $v\in W$, let $G'=G-v$, (b) If $v\not \in W$, then the single neighbor $u$ of $v$ in $G$ is in a minimum $W$-dominating set of $G$, and instance $\{G,W\}$ reduces to $\{G',W'\}$ where $G'= G - \{u,v\}$, and $W'$ results by dropping all neighbors of $u$ from $W$. The output of the leaf reduction operation is $\{G',W'\}$ and (in case-b) vertex $u$. 
\end{lemma}
\begin{proof}
  Case (a) follows by definition. In case (b), if $S$ is a minimum $W$-dominating set of $G$ that contains $v$, then $S = S-v \cup u$ is also a minimum dominating $W$-dominating set. Any dominating set must contain either $u$ or $v$ so that $v$ is dominated, and thus we can consider $u$ as part of the dominating set, and remove it from $G$. By definition, the  neighbors of $u$ do not need be dominated in $G'$.  Also, $v$ cannot dominate any vertices, so dropping from $G$ is correct.  
\end{proof}

\begin{lemma} [Dangling Path Domination]
 \label{lem:danglingPath}
 Let $\{G, W\}$ be an instance of the {\sc RDS} problem. Suppose $G$ contains a path $P$ of the form $u-q-v$, where $q$ is a path and $u$ is a leaf. Then in linear time we can reduce $\{G, W\}$ to an instance $\{G', W'\}$, where $G'$ does not contain the vertices in $u-q$. The output of the operation is $\{G',W'\}$ and a set $S$ that dominates all vertices in $G-G'$ and is contained in a minimum $W$-dominating set of $G$.
\end{lemma}
\begin{proof}
Apply Leaf Reduction operations (Lemma~\ref{lem:leaf}) iteratively until all vertices of $u-q$ are eliminated, and let $\{G',W'\}$ be the final output of this process. During this process let $S$ be the set that contains all $u$-vertices output by the Leaf Reduction operations during the process. If $S$ contains $v$, then we drop $v$ from $G'$ and add all its neighbors in $W'$. The proof follows from an easy inductive argument. 
\end{proof}

\begin{lemma} [Path  Relaxed Domination] \label{lem:path}
Suppose $\{G, W\}$ is an instance of the {\sc RDS} problem where $G$ is a path graph. A minimum $W$-dominating set for $G$ can be found in linear time.
\end{lemma}
\begin{proof}
Apply Leaf Elimination operations (Lemma~\ref{lem:leaf}) iteratively until $P$ is exhausted. 
\end{proof}

\begin{definition}[{\sc RDSC:} Relaxed Dominating Set on Cycles] \label{lem:RDSC}
We define the {sc RDSC} problem:\\
Input: A cycle or path graph $C=(V,E)$. Two sets $U,W \subseteq V$, where $U\cap W = \emptyset$. A special vertex $t$. \\
Output: \\
Case-1: A minimum $W$-dominating set $S$ for $C$  under the  constraint that $U\subseteq S$ and $t\in S$.  \\
Case-2: A minimum $\{W\cup t\}$-dominating set $S$ for $C$, under the constraint that $U\subseteq S$ and $t\not \in S.$
\end{definition}

\begin{lemma}[Cycle Relaxed Domination]
The RDSC problem can be solved in linear time. 
\end{lemma}

\begin{proof}  Set $U$ includes vertices that will be added to the under-construction dominating set $S$. In case-1 we add~$t$ to $U$. 
Set $W$ includes the vertices for which there is no requirement to be dominated. We also add to $W$ all vertices that are neighbors of vertices in $U$. We further introduce a set of vertices $X\subseteq V$ intended to denote the vertices that are not candidates for inclusion in $S$. Initially we set $X=U$, and in case-2 we add $t$ to $X$ and to $W$. Notice that $X$ always contains at least one vertex. 

We start by traversing $C$ once. Whenever we find an edge $(u,v)$ such that both $u$ and $v$ are in $W\cup U$ then we add both $u$ and $v$ to $X$. That means that $u$ and $v$ are not candidates for inclusion in $S$, and this is due to the fact that they can  dominate only one vertex that is not already dominated, and thus there is a minimum $S$ not including them. We then drop from $C$ all vertices in $X$, as we know that there exists a minimum dominating set not containing $X$. That leaves a forest $F$ of paths, including trivial singleton vertices. These paths may contain vertices from $W$. We now need to dominate minimally each path, and we can do that in linear time by invoking the algorithm from Lemma~\ref{lem:path}. We add the dominating set for each path to $S$. Finally, we include in $S$ all vertices of $U$. 
\end{proof}

\subsubsection{Path Reduction Lemmas}
\label{sec:path-reduction}

We now give some Lemmas that reduce induced paths in an instance $\{G,W\}$ of the {\sc RDS} problem.

\begin{definition} [Induced Path] \label{def:induced}
   Let $G=(V,E)$ be a graph, and $p=(V_p,E_p)$ a path induced by its vertices in $G$. The length of $p$ is defined as the number of its edges. 
\end{definition}   

\begin{lemma} [Long $W$-free path reduction] Let $\{G,W\}$ be an instance of the {\sc RDS} problem. Suppose $G$ contains a path $p= (u,u_1,u_2,u_3,v)$ such that $u_1,u_2,u_2 \not \in W$. Let $\{G',W\}$ the same problem where $p$ has been replaced by the edge $(u,v)$ in $G'$. Then $\{G,W\}$ reduces to $\{G',W\}$. \label{lem:wfree}
\end{lemma}
\begin{proof}
Consider an arbitrary minimum $W$-dominating set $S$ for $G$, and let $|S|=t$. If at least one of $u,v$ is in $S$, then $S$ must contain one vertex from $u_1,u_2,u_3$. There is also the case that $S$ dominates all vertices in $p$ by using $u_1,u_3$. However, $S-u_1\cup u$ is also a minimum dominating set. Thus we can assume that $S$ always contains exactly one vertex from $u_1,u_2,u_3$, and that dropping that vertex from it gives a dominating set of $G'$ of size $t-1$
On the other hand, any $W$-dominating set $S'$ of $G'$ can be transformed into a dominating set of $G$, by adding vertex $u_2$.

These facts imply that if the size of the minimum $W$-dominating set in $G$ is $m$ then $G'$ contains a dominating set of size $m-1$, and it cannot contain a smaller one. Thus, computing a minimum $W$-dominating set in $G'$ guarantees that we  will find a set of size $m-1$ that can then be converted to a size $m$ dominating set for $G$, by adding $u_2$. 
\end{proof}

\begin{lemma} [Induced Path Reduction] \label{lem:pathReduction}
Let $\{G,W\}$ be an instance of the {\sc RDS} problem. The instance can be reduced in linear time to another instance $\{G',W'\}$ where $G'$ contains induced paths of length at most 9. 
\end{lemma}
\begin{proof}
The collection of all induced paths in $G$ can be found in linear time. Consider an induced path $p$ with endpoints $w_s,w_t$.

We first scan $p$ and apply Long $W$-free Path elimination (Lemma~\ref{lem:wfree}) as many times as possible, in linear time. Notice then that $w_s$ is at most 2 vertices away from a $W$-vertex of $w_t$; the same applies symmetrically for $w_t$.

We then scan $p$ once. Suppose $p$ contains a segment of the form $w_1-w_2$, where $w_1,w_2 \in W$.
Recall that both $w_1,w_2$ are already dominated. Since the edge $(w_1,w_2)$ is not needed in the minimum dominating set, we can drop it. This disconnects $p$ into two `dangling' paths $w_s-\ldots-w_1$ and $w_2-\ldots-w_t$ that can be further reduced using Dangling Path Domination (Lemma~\ref{lem:danglingPath}).

Notice now that if $p$ has not been already eliminated then any consecutive $W$-vertices on $p$ must be separated by either 1 or 2 vertices that are not in $W$. We then scan the path from $w_s$ to $w_t$ and find the first three vertices $w_1,w_2,w_3 \in W$. If no such vertices exist, then we stop. Otherwise, we have the following cases:

\smallskip
\noindent \textbf{Case (b-1):} If we have a path segment of the form $w_1-u-w_2-v-w_3$, where $u,v \not \in W$, then we replace it with $w_1 - u -v - w_3$ to get $G'$ and let $W'=W-w_2$. We claim that if $\{G,W\}$ has a minimum dominating set of size $m$, then $\{G',W'\}$ has a dominating set of size $m$. And if $\{G',W'\}$ has a dominating set of size $t$, then $\{G,W\}$ has a dominating set of size $t$. Together, these imply that $\{G,W\}$ reduces to $\{G',W'\}$.

For the proof, let $S$ be an arbitrary minimum $W$-dominating set in $G$.
We have the following cases: (i) If $w_1,w_3 \in S$ then segment $q=u-w_2-v$ is already dominated (since $w_2$ does not have to be dominated). We can also see that $S$ dominates all vertices in $G'$. (ii) If $w_1\in S$ and $w_3 \not \in S$, then it must be the case that $v\in S$. We then see that $S$ is also a dominating set in $G'$. (iii) The same applies symmetrically when $w_3\in S$ and $w_1 \not \in S$. (iv) If $w_1,w_3 \not \in S$, then it must be the case that $w_2 \in S$. In that case, we let $S' = S-w_2\cup u$, and get the a dominating set of the same size in $G'$. In the opposite direction, let $S'$ be a $W'$-dominating set in $G'$. (i) If $w_1,w_3 \in S'$, then $S=S'$ is a dominating set in $G$. (ii) If $w_1\in S$ and $w_3\not \in S$, then  it must be that $u\in S'$ or $v\in S'$, and setting $S= S'\cup w_2-\{u, v\}$ gives a same-size dominating set in $G$.  (iii) If $w_3\in S$ and $w_1\not \in S$, similarly with the previous case. (iv) Finally, when $w_1,w_3\not \in S$, then without loss of generality we have $u\in S'$. Then $S = S'\cup w_2 - u $ is a dominating set for $G$. %Overall, a size-$m$ minimum $W$-dominating set in $G$ can be converted into size-$m$ $W'$-dominating set in $G'$, and a size-$t$ $W'$-dominating set in $G'$ can be converted into a size-$t$ $W$-dominating set in $G$. Thus, computing a minimum $W'$-dominating set in $G'$ will find a size-$m$ dominating set that can then be converted to a minimum $W$-dominating set in $G$. 

\smallskip
\noindent \textbf{Case (b-2):} If we have a path segment  of the form $q = w_1-u_1-u_2-w_2-v_1-v_2-w_3$, where $u_1,u_2,v_1,v_2 \not \in W$, then we can replace that with  $q' = w_1-u-v-w_3$, and let $W'= W-w_2$.
We claim that if $\{G,W\}$ has a minimum dominating set of size $m$, then $\{G',W'\}$ has a  dominating set of size $m-1$. And if $\{G',W'\}$ has a dominating set of size $t$, then $\{G,W\}$ has a dominating set of size $t+1$. Together, these imply that $\{G,W\}$ reduces to $\{G',W'\}$.

For the proof, let $S$ be an arbitrary minimum $(U,W)$-dominating set of $G$. We have the following cases: (i) If $w_1,w_3 \in S$, 
then $w_2\in S$. Setting $S'=S-{w_2}$ dominates $G'$. We also note that if $w_1\in S$, then applying on the segment the algorithm of Lemma~\ref{lem:danglingPath}, we get that $w_3\in S$. So, the only remaining case is when $w_1,w_3 \not \in S$. In that case, two internal vertices from $q$ are in $S$, and wlog assume these are $u_2,v_2$. Then, $S'=S-\{u_2,v_2\}\cup u$ dominates $G'$. In the opposite direction, let $S'$ be a $W'$-dominating set in $G'$. As in the case of $q$, if we assume that $w_1\in S'$, then we can assume wlog that $w_3\in S'$. Then $S= S'\cup w_2$ is a dominating set for $G$. If $w_1,w_3 \not \in S$, then one of $u,v$ must be in $S'$, and assume wlog that is $u$. Then $S = S' - u\cup \{u_2,v_2\}$ is a dominating set for $G'$. 

\smallskip
\noindent \textbf{Case (b-3):} If we have a path of the form $q = w_1-u_1-u_2-w_2-v_1-w_3$, where $u_1,u_2,v_1 \not \in W$. Then the path can be replaced by $w_1 - u - w_3$.  We let $W'= W-w_2$. We claim that if $\{G,W\}$ has a minimum dominating set of size $m$, then $\{G',W'\}$ has a  dominating set of size $m-1$. And if $\{G',W'\}$ has a dominating set of size $t$, then $\{G,W\}$ has a dominating set of size $t+1$. Together, these imply that $\{G,W\}$ reduces to $\{G',W'\}$.

For the proof consider an arbitrary minimum dominating set $S$ for $\{G,W\}$. If $w_1\in S$, then applying the Dangling Path Domination algorithm, (Lemma~\ref{lem:danglingPath}) on $q$ implies that we can assume without loss of generality that $w_3\not \in S$, and $w_2\in S$. Then $S'= S - w_2$ is a dominating set for $\{G',W'\}$. 
If $w_1 \not \in S$, then again by the Dangling Path Domination algorithm we get that wlog we have $u_2, w_3\in S$.Then, $S'=S-u_2$ is a dominating set for $\{G',W'\}$. In the opposite direction, let $S'$ be a dominating set for $\{G',W'\}$. If $S'$ contains both $w_1,w_3$, $S = S'\cup u_2$ is a dominating set for $\{G,W\}$. If $w_1 \in S'$ and $w_3 \not \in S'$, then $S = S'\cup w_2$ 
is a dominating set for $\{G,W\}$. If $w_3\in S'$ and $w_1 \not \in S'$, then $S = S'\cup u_2$ is a dominating set for $\{G,W\}$. Finally, if $w_1,w_3 \not \in S'$, then $u\in S'$ and $S = S'\cup \{u_1,v_1\}-u$ is  a dominating set for $\{G,W\}$. 

We can thus work iteratively on the path, applying Cases (b), until $p$
contains at most 2 internal $W$-vertices. Besides these $W$-vertices, $p$ can have another 6 internal vertices not in $W$, due to the previous elimination of long W-free paths. The lemma follows by a simple inductive argument. 
\end{proof}

\subsubsection{Dangling Cactus Optimal Domination} \label{sec:dangling}

\begin{lemma}[Optimal Domination of Vertex-attached Cactus]\label{lem:optimal}
 Let $G = (V_G,E_G)$ be a graph and $H = (V_H,E_H)$ be a cactus with root $u$, such that $V_G \cap V_H = u$. Then a minimum dominating set $S_H$ for $V_H-u$ that possibly contains $u$ and being a subset of a minimum dominating set for $G$ can be computed in linear time. Furthermore, if $V_H-u$ has a minimum dominating set containing $u$, then $u$ is in $S_H$.
\end{lemma}
\begin{proof}
  Let $T_c = (V_c,E_c)$ be the rooted tree structure of $H$.  % Let $below(v)$ denote the set of vertices appearing before $v$ in a postorder traversal of $T_c$.  
  We will describe an algorithm that iteratively builds $S_H$ by processing the vertices of $T_c$ in a postorder traversal. We denote by $W$ the set of vertices that are dominated by $S_H$; every time $S_H$ is updated $W$ is also updated, implicitly. Let $B(v)$ be the set of vertices appearing before $v$ in the postorder.  Let $a(v)$ be the vertex appearing after $v$ in the postorder; in the case $v$ is the root of $T_c$, let $a(v)=nil$.
  
  \smallskip
  
  {\em Initialization:} We set $v$ to be the first vertex in the postorder traversal and set $S_H=\emptyset$.
  
  \smallskip
  
  {\em Inductive Properties:} We let $S_v$ denote the contents of $S_H$ just before processing vertex $v\in V_c$. We work with two properties to be satisfied for each $v\in V_c$: \textbf{P-(a)}   For each $v' \in B(v)$, $S_v$ dominates all vertices in $C(v')$ with the possible exception of its articulation vertex $w'$ in the case when $w'\in C_v$. 
  \textbf{P-(b)}~$S_v$ is a subset of a minimum dominating set of $G$.  
  
  \smallskip
  
  We will describe the algorithm while simultaneously proving its correctness. The algorithm works in iterations. In each iteration, the algorithm processes a vertex $v$ of $T_c$. In order to prove correctness we show that after processing $v$ the two inductive properties are satisfied for $a(v) \in V_c$. To facilitate the proof we fix an arbitrary minimum dominating set of $G$ and in each iteration we update $S'$ so that $S_H \subseteq S'$, without increasing the size of $S'$, from which it will follow that P-(b) is satisfied after each step. We note that if P-(a) and P-(b) are satisfied for $S_{nil}$ then the Lemma follows.  

  \smallskip  

  {\em Algorithm:}  Before processing $v$, note that $S_v$ may already include vertices in $C(v)$ or dominate some vertices from $C(v)$. We then apply one of the following possible cases and corresponding steps: 
  
  \medskip
  \noindent \textbf{(a)}: $C(v)$ is dominated by $S_v$. In this case we take no action. 
  
  \smallskip
  \noindent {\em Correctness:}   $S_{a(u)}$ already satisfies the two properties because $C(v)$ is already dominated.  
  
  \smallskip
  \noindent \textbf{(b)}: $C(v)$ is a single vertex $v'$ not dominated by $S_v$. Let $u$ be the parent of $v$ in $T_c$ and let $(w,v')$ be the edge that connects $v'$ with $C(u).$ We then let $S_H = S_H \cup w$. 
  
  \smallskip
  \noindent {\em Correctness:} P-(a) is satisfied by construction for $S_{a(v)}$. If $v'\in S'-S_H$ then by the inductive property for $S_v$, $v'$ is needed only for dominating itself and possibly $w$, and thus we can set $S' = S'\cup w -v'$.

  \smallskip
    \noindent \textbf{(c)}: $C(v)$ is a non-trivial cycle. Let $U\subset C(v)$ be the vertices of $C(v)$ that are already included in $S_v$ and $W\subset C(v)$ be the vertices already dominated by $S_v$. Let $t$ be the articulation vertex of $C(v)$. If $C(v)$ is edge-articulated let $w\in V_G$ be the vertex to which $v$ is connected to the parent cycle of $C(v)$.  We solve the {\sc RDSC} problem (lemma~\ref{lem:RDSC}) twice  with inputs $C(v),U,W,t$, which gives two dominating sets $S_1$ (Output Case-1) and $S_2$ (Output Case-2). In the case when $t\in S_H$, we let $S_H = S_H \cup S_1$. Otherwise, because $S_1$ dominates one extra vertex relative to $S_2$, we have $|S_2| \leq |S_1|$. If $|S_2|=|S_1|$ we let $S_H = S_H \cup S_1$. If $|S_2|<|S_1|$, we let $S_H = S_H \cup S_2$, and if $C(v)$ is edge-articulated, we let $S_H = S_H \cup w$. 
    
   \smallskip
   \noindent {\em Correctness:}  Independently of whether $S_1$ or $S_2$ has been used to augment $S_H$, P-(a) is satisfied by construction for $a(v)$.    Let $u \in C(v)$, such that $u \neq t$ and $u\not \in S_V \cup W$. Because $u\neq t$, it can be dominated only by vertices in $C(v)$, or vertices that belong to cycles in $B(v)$.  Suppose that $u\in C(v)$ is dominated in $S'$ by a vertex $u'$ in $B(v)$. Then $u'$ is needed only for dominating $u$; this is because $u\not \in W$ implies that its neighbors in $B(v)$ are already dominated by $S_v$. Thus we can set $S' = S' - u'+u$. This implies that we can update $S'$ so that all vertices in $C(v)-t$ are dominated by the vertices in $S_v$ and a number $k$ of a additional vertices from $C(v)$. 
   
   If $t\in S_v$, or in the case when $|S_1|=|S_2|$ the number $k$ of additional vertices is $|S_1|$ which implies that $S'-S_v$ contains a set $S_1'$ with exactly $|S_1|$ vertices from $C(v)$. We can thus set $S' = S'\cup S_1 - S_1'$, which ensures that $S'$ has minimum size and $S_H \subseteq S'$.  
   
   In the case $|S_2|<|S_1|$ the number $k$ of additional vertices (excluding $t$) is $|S_2|$. Thus $S'-S_v$ includes a set $S_2'$ of $|S_2|$ vertices in $C(v)-t$. We can then set $S' = S' \cup S_2 - S_2'$ which ensures that $S'$ has minimum size and $S_H \subseteq S'$. In the case $C(v)$ is edge-articulated, then $t$ is not included in the parent cycle of $C(v)$. If $S'$ includes $t$ then $t$ is used to dominate only $t$ and its parent $w$, which means that $S' = S\cup w - t$ is also minimum. 
   
   Finally note that when processing $u_c$ (the root of the cactus), if its articulation vertex $u$ is not already in $S_H$, we include $u$ in $S_H$ if $S_1$ is of minimum size as a dominating set of the vertices that remain non-dominated. Thus, if there is a minimum dominating set for $V_H-u$ that includes $u$, then $u$ is included in $S_H$.
\end{proof}

\subsubsection{Completing the Proof}

\begin{theorem}
 Let $G$ be a graph that has edge feedback set number equal to $k$. Then, in linear time, the minimum dominating set problem on $G$
 can be reduced to an  instance $\{G',W'\}$, where $G'$ has at most $27k$ edges, at most $26k$ vertices and at most $2k$ vertices of degree greater than 2. 
\end{theorem}
\begin{proof}

Let $G_c,G_k$ be the graphs described in Claim~\ref{claim:decomposition}, $G$ contains a forests of cactus graphs and by Claim~\ref{claim:tree_structure} we can find the decomposition and the tree structure of the cactus graphs in linear time. 
   
Let $S'$ be an arbitrary minimum dominating set $G$. Consider a cactus $H$ in the forest. Let $S_H$ be the dominating set for $V_H-u$, where $u$ is the root of the cactus, computed by the algorithm in Lemma~\ref{lem:optimal}. Let $S_H' = S' \cap (V_H-u)$.
There are two cases:
(i) If $S_H'$ dominates $V_H-u$, then by Lemma~\ref{lem:optimal}, we must have $|S_H| = |S_H'|$, so we can set $S' = S\cup S_H - S_H'$, without increasing the size of $S'$.
(ii) If $S_H'$ does not dominate $V_H-u$, then $u$ must be in $S'$. If $V_H-u$ can be dominated by a set of less than $|S'|+1$ vertices not including $u$, then $S_H$ is such a set, and thus we can set $S' = S'\cup S_H-S_H'$ without increasing its size. Otherwise if there is a minimum dominating set for $V_H-u$ that includes $u$, then $S_H$ such a set (by Lemma~\ref{lem:optimal}), and thus $S_H-u$ can replace $S_H'$ in $S'$ without increasing its size. 
In either case, we can update $S'$ so that it contains $S_H$, without increasing its size.    We can thus compute all dominating sets of the cactus graphs $G_c$ by using the Algorithm in Lemma~\ref{lem:optimal}. Let $S$ be the union of these dominating sets. Let $W$ be the set of vertices dominated by $S$, not including $S$.
Notice that both $S$ and $W$ can intersect with $G_k$ (on the articulation vertices of the cactus graphs). We thus mark the $S$-vertices and $W$-vertices of $G_k$. We then find the neighbors of the $S$-vertices in $G_k$, add them to the set $W$, and delete the $S$-vertices from $G_k$. This reduces the original instance to a $\{G_k, W\}$ instance. We then apply Long Path Reduction (Lemma~\ref{lem:pathReduction}) on the edge-disjoint paths $\{G_k, W\}$ and further reduce the graph to an instance $\{G', W'\}$. Then each such such path in $G_k$ is replaced by at most 9 edges, and 8 vertices. The Theorem follows from the fact that $G_k$ consists of $2k$ vertices of degree greater than 2, and at most $3k$ edge-disjoint paths. 
\end{proof}

\section{Parameter: Solution Size}\label{sec:solsize}

In this section, we develop a simple FPT-approximation algorithm for {\sc Dominating Set} parameterized by $k$, the sought size of a solution. To this end, we will use the following result.

\begin{proposition}[\cite{vazirani}]\label{prop:approxDomSet}
{\sc Weighted Dominating Set} admits a (polynomial-time) $(\ln n - \ln\ln n +O(1))$-approximation algorithm.
\end{proposition}

We now turn to present the algorithm.

\begin{theorem}\label{thm:solSize}
Fix $0\leq \alpha<1$. Then, the {\sc Dominating Set} problem parameterized by the solution size $k$ admits a  $((1-\alpha)\ln n + O(1))$-approximation $n^{\alpha k+O(1)}$-time algorithm.
\end{theorem}

\begin{proof}
We first describe the algorithm. Let $(G,k)$ be an instance of {\sc Dominating Set} parameterized by the solution size $k$. For every subset $U\subseteq V(G)$ of size at most $\alpha k+1$:
\begin{enumerate}
\item Let $G_U$ be the graph where $V(G_U)=V(G-U)\cup\{x\}$ for some new vertex $x\notin V(G)$, and $E(G_U)=E(G-U)\cup \{\{x,v\}: v\in N_G(U)\}$.
\item Call the algorithm of Proposition \ref{prop:approxDomSet} with $G_U$ as input, and let $S'_U$ be its output.
\item Let $S_U=(S'_U\setminus\{x\})\cup U$.
\end{enumerate}
Return the set $S$ of minimum-size among the sets in $\{S_U: U\subseteq V(G)\}$.

Clearly, the algorithm runs in ${n\choose \alpha k+1}\cdot n^{O(1)}\leq n^{\alpha k+O(1)}$ time. For correctness, first note that, clearly, the algorithm returns a dominating set of $G$. Now, let $S^\star$ be a dominating set of $G$ of minimum-size. We need to prove that, if $|S^\star|\leq k$, then $|S|\leq ((1-\alpha)\ln n + O(1))k$. To this end, let $U^\star$ be some subset of $S^\star$ of size $\min\{\alpha k+1,|S^\star|\}$.  Consider the iteration where $U=U^\star$. Then, observe that $(S^\star\setminus U)\cup\{x\}$ is a dominating set of $G_U$, and its size is equal to $|S^\star|-|U|+1\leq (1-\alpha)k$. Hence, the call to the algorithm of Proposition \ref{prop:approxDomSet} with $G_U$ returns $S'_U$ of size bounded from above by $(\ln n - \ln\ln n +O(1))(1-\alpha)k\leq ((1-\alpha)\ln n+O(1)) k$. Then, $|S_U|\leq |S'_U|+|U|\leq ((1-\alpha)\ln n+O(1)) k + \alpha k\leq ((1-\alpha)\ln n+O(1)) k$. This completes the proof.
\end{proof}

%%%%%%%%%%%%%%%%%%%% CONCLUSION
\section{Conclusion and Future Directions}\label{sec:conclusion}

We presented algorithmic results that sidestep time complexity barriers for {\sc Dominating Set}. For this purpose, we incorporated  approximation for the parameters solution size and treewidth, larger parameterization for the parameters vertex cover and feedback edge set compared to treewidth, or both for the parameter vertex modulator to constant treewidth compared to treewidth.  

\subparagraph*{Extension of Our Approaches.} While we have focused on {\sc Dominating Set}, we believe that some of our approaches might be applicable to other problems as well. For example, consider the {\sc Graph Coloring} problem, where, given a graph $G$ and an integer $q\geq 3$, the objective is to determine whether $G$ admits a proper coloring in $q$ colors. Under the SETH, {\sc Graph Coloring} cannot be solved in  $(q-\epsilon)^{\mathsf{tw}}\cdot n^{O(1)}$ time for any fixed $\epsilon>0$~\cite{lokshtanov.known.2018}. Then, we follow a simplification of the approach we presented in Section \ref{sec:consttw}. Briefly, the idea is to consider two problems: one problem concerns the graph induced by the modulator, and the other problem concerns the rest of the graph.  So, suppose we are given a subset $M\subseteq V(G)$  such that the treewidth of $G-M$ is at most $d$ (where $d$ is a fixed constant). On the one hand, we solve {\sc Graph Coloring} on $G[M]$ in $2^{|M|}\cdot|M|^{O(1)}$ time using the algorithm in \cite{DBLP:conf/focs/Koivisto06}, and on the other hand, we solve {\sc Graph Coloring} on $G-M$ in $n^{O(1)}$ time based on straightforward dynamic programming. We consider the color sets used by the two solutions to be disjoint, thereby obtaining a $2$-approximate solution in  $2^{\mathsf{tw}_d}\cdot n^{O(1)}$ time. Essentially the same approach works for {\sc Independent Set} as well, where, given a graph $G$ and a non-negative integer $k$, the objective is to determine whether $G$ admits an independent set of size at least $k$. Under the SETH, {\sc Independent Set} cannot be solved in  $(2-\epsilon)^{\mathsf{tw}}\cdot n^{O(1)}$ time for any fixed $\epsilon>0$~\cite{lokshtanov.known.2018}.  On the one hand, we solve {\sc Independent Set} on $G[M]$ in $1.19997^{|M|}\cdot|M|^{O(1)}$ time using the algorithm in \cite{DBLP:journals/iandc/XiaoN17}, and on the other hand, we solve {\sc Independent Set} on $G-M$ in $n^{O(1)}$ time based on straightforward dynamic programming. We output the largest among the two solutions, thereby obtaining a $2$-approximate solution in  $1.19997^{\mathsf{tw}_d}\cdot n^{O(1)}$ time.

\subparagraph*{Other Observations.} We remark that there exists a simple reduction from {\sc Dominating Set} on general graphs to {\sc Dominating Set} on chordal (and even split) graphs that is approximation preserving, keeps $k$ the same and makes $n$ twice as large.\footnote{See https://cstheory.stackexchange.com/questions/47730/best-approximations-of-minimum-dominating-sets-in-chordal-graphs.}
So, under the SETH, the {\sc Dominating Set} problem on chordal (and even split) graphs does not admit a $g(k)$-approximation $f(k)\cdot n^{k-\epsilon}$-time algorithm for any computable functions $g$ and $f$ of $k$ and fixed $\epsilon>0$.

\subparagraph*{Directions for Future Research.} Firstly, we find the questions of  improvements of the performance of our algorithms (in terms of running times and approximation ratios) interesting. In particular, does {\sc Dominating Set} admit a $2$-approximation $O(2^{\mathsf{tw}}\cdot n)$-time algorithm, or a $(1+\epsilon)$-approximation $O(2^{\mathsf{tw}}\cdot n)$-time algorithm for any fixed $\epsilon>0$? Additionally, we have the following questions regarding {\sc Dominating Set}:
\begin{enumerate}
    \item Prove or refute the following conjecture:
    \begin{conjecture}\label{conj:constTwLower}
Under the SETH, there exists a fixed constant $d\in\mathbb{N}$ such that {\sc (Weighted) Dominating Set} cannot be solved in $(3-\epsilon)^{\mathsf{tw}_d}\cdot n^{f(d)}$ time for any fixed constant $\epsilon>0$ and function $f$ of $d$, where $\mathsf{tw}_d$ is the minimum size of a vertex set whose deletion from $G$ results in a graph of treewidth at most $d$.
\end{conjecture}
    \item Study {\sc Dominating Set} parameterized by the solution size plus the distance (e.g., number of vertex or edge deletions or contractions) to graph classes where it belongs to FPT, particularly planar graphs and claw-free graphs.
    \item Conduct a similar study for problems beyond {\sc Dominating Set}. Here, possibly and as argued above, the ideas presented in this article can be re-used.
\end{enumerate}

\bibliographystyle{plainurl}
\bibliography{references}
\end{document}